\documentclass[fleqn,usenatbib]{mnras}

\usepackage[T1]{fontenc}
\usepackage{url}
\DeclareRobustCommand{\VAN}[3]{#2}
\let\VANthebibliography\thebibliography
\def\thebibliography{\DeclareRobustCommand{\VAN}[3]{##3}\VANthebibliography}

\usepackage{graphicx}	
\usepackage{amsmath}	
\usepackage{amssymb}	
\usepackage{multirow}
\usepackage{booktabs}

\usepackage{ulem}

\title[ESAP-LAMOST]{Estimating Atmospheric Parameters from LAMOST Low-Resolution Spectra with Low SNR}

\author[X. Li et al.]{
Xiangru Li$^{1}$\thanks{E-mail: xiangru.li@gmail.com},
Si Zeng$^{2}$,
Zhu Wang$^{2}$,
Bing Du$^{3}$,
Xiao Kong$^{3}$,
and Caixiu Liao$^{2}$
\\
$^{1}$School of Computer Science, South China Normal University, No. 55 West of Yat-sen Avenue, Guangzhou 510631, China\\
$^{2}$School of Mathematical Sciences, South China Normal University, No. 55 West of Yat-sen Avenue, Guangzhou 510631, China\\
$^{3}$Key Laboratory of Optical Astronomy, National Astronomical Observatories, Chinese Academy of Sciences, Beijing 100012, China\\
}

\date{Accepted XXX. Received YYY; in original form ZZZ}

\pubyear{2022}

\begin{document}
\label{firstpage}
\pagerange{\pageref{firstpage}--\pageref{lastpage}}
\maketitle

\begin{abstract}
Large Sky Area Multi-Object Fiber Spectroscopic Telescope (LAMOST) acquired tens of millions of low-resolution stellar spectra. The large amount of the spectra result in the urgency to explore automatic atmospheric parameter estimation methods. There are lots of LAMOST spectra with low signal-to-noise ratios (SNR), which result in a sharp degradation on the accuracy of their estimations. Therefore, it is necessary to explore better estimation methods for low-SNR spectra. This paper proposed a neural network-based scheme to deliver atmospheric parameters, LASSO-MLPNet. Firstly, we adopt a polynomial fitting method to obtain pseudo-continuum and remove it. Then, some parameter-sensitive features in the existence of high noises were detected using Least Absolute Shrinkage and Selection Operator (LASSO). Finally, LASSO-MLPNet used a Multilayer Perceptron network (MLPNet) to estimate atmospheric parameters $T_{\mathrm{eff}}$, log $g$ and [Fe/H]. The effectiveness of the LASSO-MLPNet was evaluated on some LAMOST stellar spectra of the common star between APOGEE (The Apache Point Observatory Galactic Evolution Experiment) and LAMOST. it is shown that the estimation accuracy is significantly improved on the stellar spectra with $10<\mathrm{SNR}\leq80$. Especially, LASSO-MLPNet reduces the mean absolute error (MAE) of the estimation of $T_{\mathrm{eff}}$, log $g$ and [Fe/H] from (144.59 K, 0.236 dex, 0.108 dex) (LASP) to (90.29 K, 0.152 dex, 0.064 dex) (LASSO-MLPNet) on the stellar spectra with $10<\mathrm{SNR}\leq20$. To facilitate reference, we release the estimates of the LASSO-MLPNet from more than 4.82 million stellar spectra with $10<\mathrm{SNR}\leq80$ and 3500 < SNR$g$ $\leq$ 6500 as a value-added output.

\end{abstract}

\begin{keywords}
methods: data analysis –- methods: statistical -- stars: abundances -- stars: fundamental parameters.
\end{keywords}



\section{Introduction}

The stellar spectra contain a lot of basic information about stars, e.g., effective temperature ($T_{\mathrm{eff}}$), surface gravity (log $g$), the metallicity ([Fe/H]) and so on \citep{liu2014}. These information is a basis for researches on the stellar evolution and the history of the Milky Way. Therefore, many large-scale survey telescopes have been built, a series of spectral surveys are carried out, and the amount of stellar spectra is greatly increased. Therefore, it is urgent to explore automatic atmospheric parameter estimation methods.

Large Sky Area Multi-Object Fiber Spectroscopic Telescope \citep[LAMOST;][]{cui2012, liu2015} is a major scientific and technological infrastructure, which can simultaneously collect the spectra from up to 4000 targets. After several years of sky surveys, LAMOST has obtained more than 10 millions stellar spectra. The stellar parameters of these spectra are derived using LAMOST Stellar Parameter Pipelines \citep[LASP;][]{luo2015, wu2011automatic}. The LASP is designed based on the University of Lyon Spectroscopic analysis Software \citet[ULYSS;][]{koleva2009ulyss}. The LASP delivers stellar parameters by minimizing the $\chi^2$ between observed spectra and template spectra from ELODIE library \citep{prugniel2001database, prugniel2007new}. However, the ELODIE library lacks spectral samples of K giants and subgiants, which results in some deviation in the LASP parameter estimations from their theoretical values. On the low-resolution stellar spectra, especially, the metal lines are relatively weak and interfered by noise easily. Therefore, the accuracy of the estimated parameters from the LAMOST low-resolution stellar spectra is relatively low and needs to be improved.

\begin{figure*}
	\includegraphics[width=1.00\textwidth]{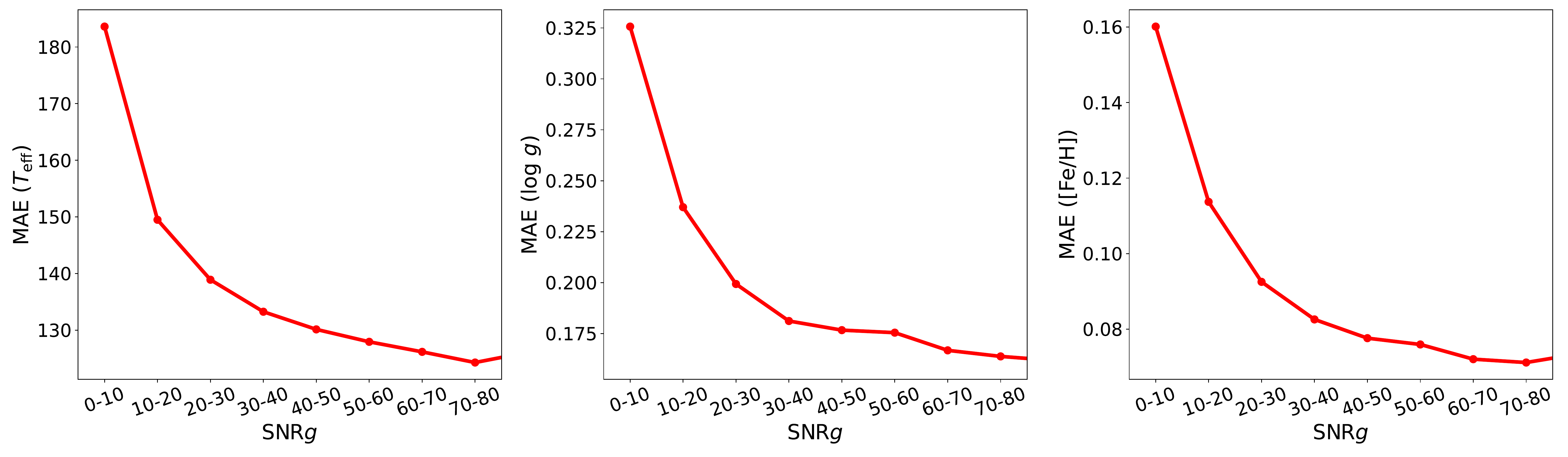}
	
	\caption{The dependencies of the error of the stellar atmospheric parameters ($T_{\mathrm{eff}}$, log $g$, and [Fe/H]) on SNR$g$. The spectral data are from the 115,942 common stars between LAMOST and APOGEE. The error refers to the mean absolute errors (MAE) between the APOGEE\_$payne$ labels and the LASP estimations.}
	\label{fig:MAE_SNRg}
\end{figure*}

Therefore, experts have carried out a series of studies for the parameter estimation problem from LAMOST low-resolution stellar spectra. For the LAMOST spectra with the low-resolution and high signal-to-noise ratio (SNR), \cite{xiang2017estimating} proposed a stellar atmospheric physical parameters, absolute magnitude and element abundance estimation method based on Kernel Principal Component Analysis \citep[KPCA;][]{scholkopf1997kernel}. This study mainly focused on LAMOST spectra with SNR > 50. \cite{ting2017measuring} studied the estimation problem of element abundance for LAMOST spectra with SNR higher than 30 for each pixel based on a single-hidden-layer neural network model. \cite{xiang2019abundance} used The DD-Payne method to determine stellar atmosphere parameters and 16 element abundances for 6 million stars with a SNR higher than 30 for each pixel. \cite{zhang2020deriving} proposed a spectral parameter estimation scheme, SLAM,  based on support vector regression (SVR) method, and predicted the parameters with the uncertainties 50 K, 0.09 dex and 0.07 dex respectively on $T_{\mathrm{eff}}$, log $g$ and [Fe/H] for the LAMOST spectra with SNR$g$ > 100. \cite{wang2020spcanet} used a Generative Spectral Network (GSN) to estimate the atmospheric physical parameters for the LAMOST stellar spectra with SNR$g$ > 30, and obtained that for the stellar spectra with SNR$g$ $\geq$ 50,  the accuracy of $T_{\mathrm{eff}}$, log $g$, [Fe/H] and [$\alpha$/Fe] is 80 K, 0.14 dex, 0.07 dex, and 0.168 dex, respectively, and these results have a good consistency with other studies. LAMOST collaboration \citep{luo2015,wu2014automatic} uses LASP -- a method based on minimizing the $\chi^2$ between the measured spectra and the synthetic spectra of the ELODIE library -- to determine the stellar spectral parameters and radial velocity for the stellar spectra with SNR$g$ > 6 and SNR$g$ > 15 for the dark and bright nights respectively \citep{luo2015}.

\begin{figure}
\centering
	\includegraphics[width=8cm]{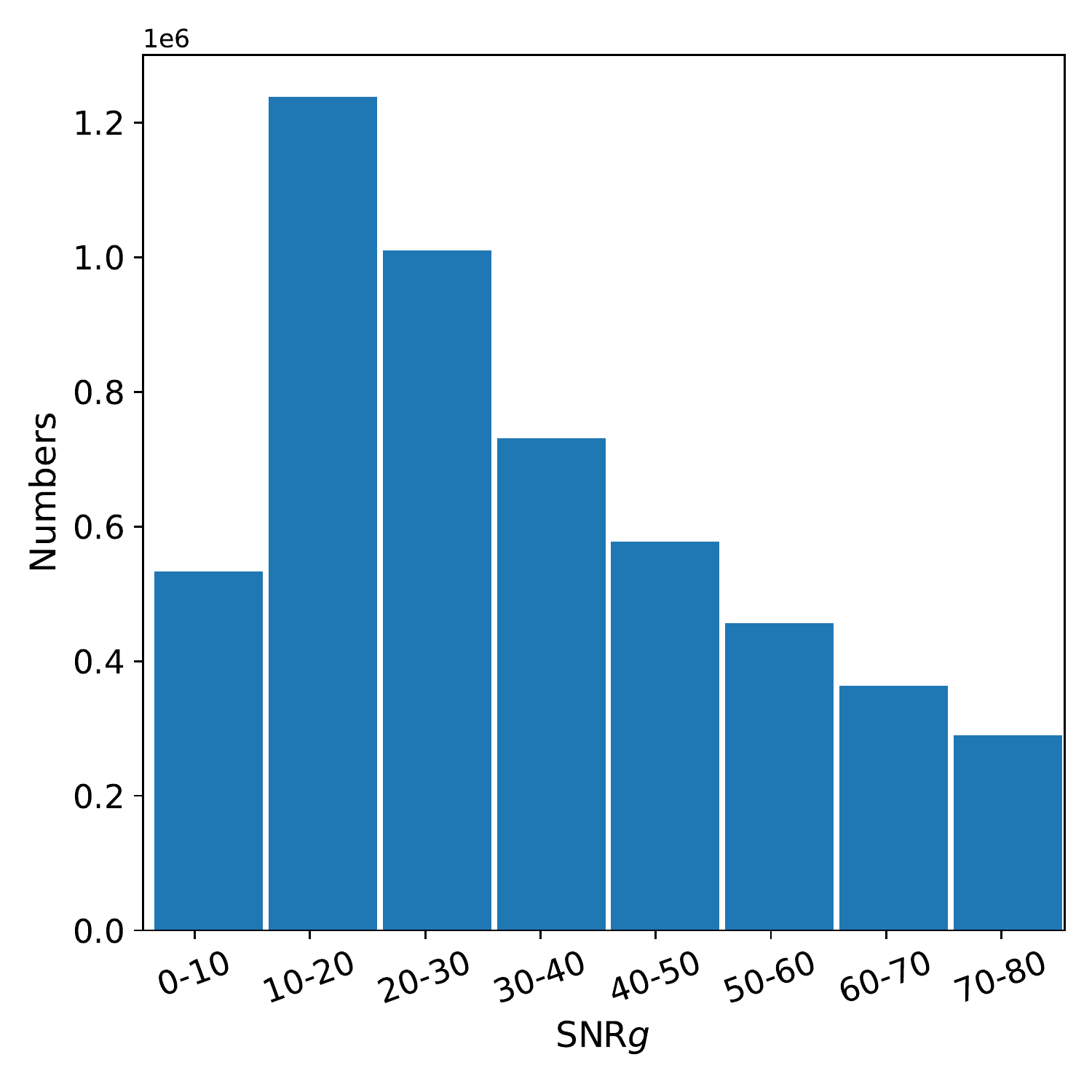}
    \caption{The distribution of the LAMOST DR8, low-resolution stellar spectra on SNR$g$.}
    \label{fig:number_distribution}
\end{figure}

The current related researches mainly focus on the parameter estimation problem of the LAMOST spectra with higher SNR and most studies used the neural network models. Although the official LAMOST data provides parameter estimation for the spectra with SNR higher than 6, its accuracy degrades sharply as the SNR decreases (Figure \ref{fig:MAE_SNRg}), and the scale of such spectra is huge, in which there are more than 5 million spectra with SNR less than 80. And there are more than two millions low-resolution stellar spectra with 10 < SNR$g$ < 30 (Figure \ref{fig:number_distribution}). These spectra are affected by a lot of noise, and their metal lines are mixed terribly. These characteristics result that the spectral physical properties are not obvious, and the accuracy of each parameter drops sharply with the decrease of SNR$g$ (Figure \ref{fig:MAE_SNRg}). Therefore, it is necessary to conduct special research on stellar spectra with low SNR to further enhance the scientific value of LAMOST observation data. In this paper, we focus on designing a machine learning scheme based on LASSO (Least Absolute Shrinkage and Selection Operator) and a neural network to improve the accuracy of atmospheric physical parameters for LAMOST low-SNR stellar spectra

In this paper, our scheme consists of three procedures. First, this paper used a polynomial fitting method to estimate the pseudo-continuum for every spectrum, and removed the pseudo-continuum. Second, we detected some parameter-sensitive features in the existence of high noises using LASSO. The LASSO evaluated the effectiveness of spectral features based on the combined effects of spectral fluxes and the noises overlapped on them. Finally, we proposed a Multilayer Perceptron neural network-based method to estimate the stellar atmospheric parameters $T_{\mathrm{eff}}$, log $g$ and [Fe/H].

This paper is organized as follows. Section~\ref{sec:data} describes the data used in this work and some data preprocessing procedures. In Section~\ref{sec:method}, we introduce the stellar atmospheric parameter estimation method LASSO-MLPNet. Section~\ref{sec:experiment} describes performance evaluation results. Section~\ref{sec:application} presents the application of our model to the LAMOST DR8 low-resolution stellar spectra. Finally, a brief summary is made in Section~\ref{sec:conclusions}.


\section{Data and Their Preprocessings}
\label{sec:data}

In stellar parameter estimation, we usually use synthetic or empirical spectra as spectral templates or training sets (reference sets) to establish a pipeline. The synthetic spectra are computed from the stellar atmospheric physical model. Therefore, the parameter coverage, and wavelength range of the synthetic spectral libraries can be set according to requirements. However, this scheme depends on the limited physical information we currently have about stellar objects, which makes a certain difference between the synthetic spectra and the observed spectra, and even makes a large biases in certain parameter ranges. These differences and biases can introduce some systematic uncertainties into the establishment of the parameter estimation model. For example, the opacity of hydrogen atoms, metal atoms and molecules, and the large discrepancy between the modeling for hot \& cold stars and the situation of actual celestial bodies, etc \citep{plez1992spherical, masseron2014ch}.

The empirical spectral libraries are generally composed of observed spectra which have been strictly and accurately parameter-calibrated. However, the empirical spectral libraries generally are limited by their numbers of samples, the narrow coverages of the parameters and wavelengths. For example, the ELODIE spectral library \citep{prugniel2007new} lacks K-giant and subgiant spectral samples, and their wavelength range is between 4000 - 4800 \r A. In particular, if the empirical spectral library and the spectra to be parameterized come from different astronomical telescopes, instrumental effects will have a significant impact on model establishment and model accuracy. The instrumental effect means that, even for the observations of the same celestial body, there is a certain biases between the observed spectra from different telescopes.

Based on the above-mentioned considerations, the reference set in this paper adopted the label transfer empirical spectral library. This kind spectral library consists of a series of LAMOST observed spectra, whose reference labels are transferred from the parameter estimation results from the high-resolution and high-SNR spectra from common sources. It is well known that the atmospheric parameter estimation results from stellar spectra with high-resolution and high-SNR are usually more accurate. For example, APOGEE \citep{majewski2017} (The Apache Point Observatory Galactic Evolution Experiment) adopts ASPCAP \citep{perez2016} (APOGEE Stellar Parameter and Chemical Abundances Pipeline) to determine the stellar spectral atmospheric physical parameters and chemical element abundances, and the parameter accuracy of giant stars is more reliable than that of dwarf stars. The three atmospheric physical parameters ($T_{\mathrm{eff}}$, log $g$ and [Fe/H]) are accurate to 2\%, 0.1 dex and 0.05 dex, respectively, and the accuracy of 15 chemical element abundances is typically under 0.1 dex \citep{perez2016}. \cite{ting2019payne} used the improved Kurucz line list to train The Payne model, and estimated atmospheric parameters and 15 element abundances for approximately 230,000 APOGEE high-resolution stellar spectra, and an APOGEE\_$payne$ catalog was released based on the estimated results. This catalog includes some giant stars and dwarf stars. The verification of that paper shows that the accuracy of its parameter estimation results is significantly improved on the basis of ASPCAP: the accuracy of $T_{\mathrm{eff}}$ is under 30 K, the accuracy of log $g$ is 0.05 dex, and the accuracy of chemical element abundances is under 0.05 dex. Therefore, this work used the LAMOST spectra from the common stars between LAMOST and APOGEE as reference spectral library. The reference parameters of these LAMOST spectra are transferred from the APOGEE\_$payne$ catalog. The essence of this scheme is to use the common stars between LAMOST and APOGEE as the calibration stars.

\subsection{LAMOST Survey and APOGEE Survey}

LAMOST \citep{zhao2012lamost} is also referred to as "Guo Shoujing Telescope" with a new type of large-aperture (1.75m) and a large field of view (5 degrees field of view). It can observe 4000 targets simultaneously in a field of view of 20 $\mathrm{deg}^2$ in the sky, and improve effectively the target spectral collection rate. LAMOST DR8 released more than 11 millions low resolution spectra, which cover the wavelength range of 3690 \r A - 9100 \r A with resolution of 1800. There are 10,388,423 stellar spectra, and the rest are the spectra of galaxies, quasars or unknown celestial bodies. At the same time, based on LASP, LAMOST determined the stellar spectral parameter ($T_{\mathrm{eff}}$, log $g$, [Fe/H] and Radial Velocity) with SNR$g$ > 6 and SNR$g$ > 15 for the dark and bright nights respectively, and provided three stellar parameter catalogs, in which the LAMOST low-resolution A, F, G and K-type catalog contain 6,478,063 stellar spectral parameter estimates. 

APOGEE is one of the projects in the Sloan Digital Sky Survey (SDSS-$\mathrm{\uppercase\expandafter{\romannumeral3}}$). The project uses a 2.5-meter Sloan telescope to complete a three-year observation campaign, observing high-resolution (R$\sim$22,500), high-SNR (>100) and near-infrared (1.51$\mu$m-1.70$\mu$m) spectra, and uses these spectra to systematically study the stellar composition of the Milky Way. APOGEE DR14 released the spectra, their atmospheric parameters and elemental abundances for approximately 260,000 stars \citep{holtzman2018}. These stellar parameters are estimated by ASPCAP \citep{perez2016}. For dwarf stars, however, log $g$ from ASPCAP does not follow the main sequence expected from the isochron \citep{jonsson2018apogee}. Therefore, \citet{ting2019payne} used The Payne algorithm to further improve the accuracy of the effective temperature, surface gravity and 15 element abundances for APOGEE stellar spectra, and released a catalog for reference, APOGEE\_$payne$. This catalog contains approximately 230,000 stellar spectra, and the parameter ranges of $T_{\mathrm{eff}}$, log $g$ and [Fe/H] are [3,050, 7950] K, [0, 5] dex and [-1.45, 0.45] dex, respectively.

\subsection{Reference Set}

This work obtained a reference set by cross-matching the low-resolution stellar spectra from LAMOST DR8 with APOGEE\_$payne$ catalog using the TOPCAT \citep[the Tool for OPerations on Catalogues and Tables;][]{taylor2017topcat}. In the TOPCAT, we limit the maximum angular interval (Max Error) to 3 arcsecs between two stellar position coordinates (Ra, Dec). The Match Selection is set that each LAMOST spectrum has the best matching parameter from APOGEE\_$payne$ catalog. we obtained 115,942 LAMOST stellar spectra with APOGEE\_$payne$ labels. Figure~\ref{fig:MAE_SNRg} shows the inconsistencies between the LASP estimation of ($T_{\mathrm{eff}}$, log $g$ and [Fe/H]) and APOGEE\_$payne$ catalog on these spectra. It is shown, when $\mathrm{SNR}g \leq 20$, the parameter estimation accuracy drops sharply with the decrease of SNR. Therefore, there is a lot of room for improvements on the estimated atmospheric parameters ($T_{\mathrm{eff}}$, log $g$ and [Fe/H]) from these Low-SNR stellar spectra.

Therefore, this paper focused on improving the parameter estimation accuracy of the spectra with $10 < \mathrm{SNR}g \leq 20$, and limited the SNR between 10 and 20 for the reference spectra. To ensure the quality of the reference labels, we further limited the label quality of APOGEE\_$payne$ to ``good'', and ended up with 9,873 spectra of common stars for this study. The ranges of the atmospheric parameters ($T_{\mathrm{eff}}$, log $g$ and [Fe/H]) are [3,692, 7405] K, [0.679, 4.991] dex and [-1.448, 0.439] dex, respectively on these spectra. We divided these spectral objects into a training set and a test set (test set 1) based on a ratio of 8:2. Finally, 7,898 spectra were used to train the model and 1975 spectra were used to test the performance of the model.

Some experiments shows that the model learned from above-mentioned data also has application advantages on a wider range of stellar spectra. Therefore, in Section~\ref{sec:LASP_APOGEE}, we used 56,198 spectra with 5 < SNR$g$ $\leq$ 10 and 20 < SNR$g$ $\leq$ 80 from the common stars  as another test set (test set 2) to further evaluate the applicability of our model on a wider range of stellar spectra. Here, the LAMOST low-resolution stellar spectra are used as the reference spectra, and the APOGEE\_$payne$ labels are used as the reference labels.

\subsection{Preprocessing the Spectra}

The observed stellar spectra suffer from extinction, reddening, stray light pollution, instrumental noises, and so on. Therefore, the metal lines of the spectrum are not obvious, and we must properly preprocess the stellar spectral data to promote the availability of the spectral characteristic information before extracting features and estimating parameters. The specific preprocessing steps are as follows:

\begin{table*}
	\centering
	\caption{The dependencies of parameter estimation performance on window width. WW: window width of the Median Filter.}
	\label{tab:different_filter}
	\begin{tabular}{lccccccccr} 
		\hline
		\multirow{2}{*}{WW}&  & $T_{\mathrm{eff}}$ &  &  & log $g$ & &  & [Fe/H] &  \\\specialrule{0em}{1.5pt}{1.5pt}
		\cline{2-10}
		\specialrule{0em}{1.5pt}{1.5pt}
		& MAE & $\mu$ & $\sigma$ & MAE & $\mu$ & $\sigma$ & MAE & $\mu$ & $\sigma$\\
		\hline
		1 & 94.07 & 9.856 & 164.1 & 0.162 & 0.014 & 0.272 & 0.068 & -0.002 & 0.100\\
		\textbf{3} & \textbf{90.29} & \textbf{-2.892} & \textbf{152.6} & \textbf{0.152} & \textbf{0.017} & \textbf{0.258} & \textbf{0.064} & \textbf{0.004} & \textbf{0.096}\\
		5 & 97.26 & -0.438 & 164.9 & 0.154 & 0.018 & 0.264 & 0.067 & 0.006 & 0.099\\
		7 & 93.37 & 5.433 & 162.6 & 0.156 & 0.011 & 0.258 & 0.067 & 0.005 & 0.097\\
		9 & 94.23 & -8.180 & 159.0 & 0.158 & 0.013 & 0.257 & 0.068 & -0.004 & 0.097\\
		\hline
	\end{tabular}
\end{table*}

\begin{itemize}
	\item Removing redshift: we shifted each spectrum to its rest frame using the redshift estimated from the LASP.
	\item Resampling: we calculated the common wavelength range 3881\r A - 8890\r A for all stellar spectra, and resampled all spectra by a linear interpolation method with a step of 0.0001 in logarithmic wavelength coordinates.
	\item Denoising spectral data: the observed spectra usually have some bad pixels or pixels polluted by noises, which may negatively affect the parameter estimation. Therefore, it is necessary to denoise the spectral data. In this paper, we used a Median Filter method to reduce spectral noise. In Median Filter, as the width of the filtering window increases, the output signal becomes smoother and smoother, which may result in some loses of effective spectral features. Therefore, the width of the filtering window should be set according to the actual situation. In this paper, we experimentally evaluated different settings with window width of 1, 3, 5, 7, etc. The experimental results show that the Median Filter with a width 3 achieves the best performance (Table~\ref{tab:different_filter}).
	\item Normalizing spectra: Due to the uncertainty of flux calibration in LAMOST stellar spectra, and the unknown extinction values for most observed targets (especially those in the galactic disk), it is necessary to normalize the stellar spectra to improve the reliability of parameter estimations. This work obtained the pseudo-continuum by iteratively fitting a sixth-order polynomial, and normalize the spectrum by dividing its fluxes by the pseudo-continuum.
\end{itemize}

\begin{figure*}
	\includegraphics[width=0.65\textwidth]{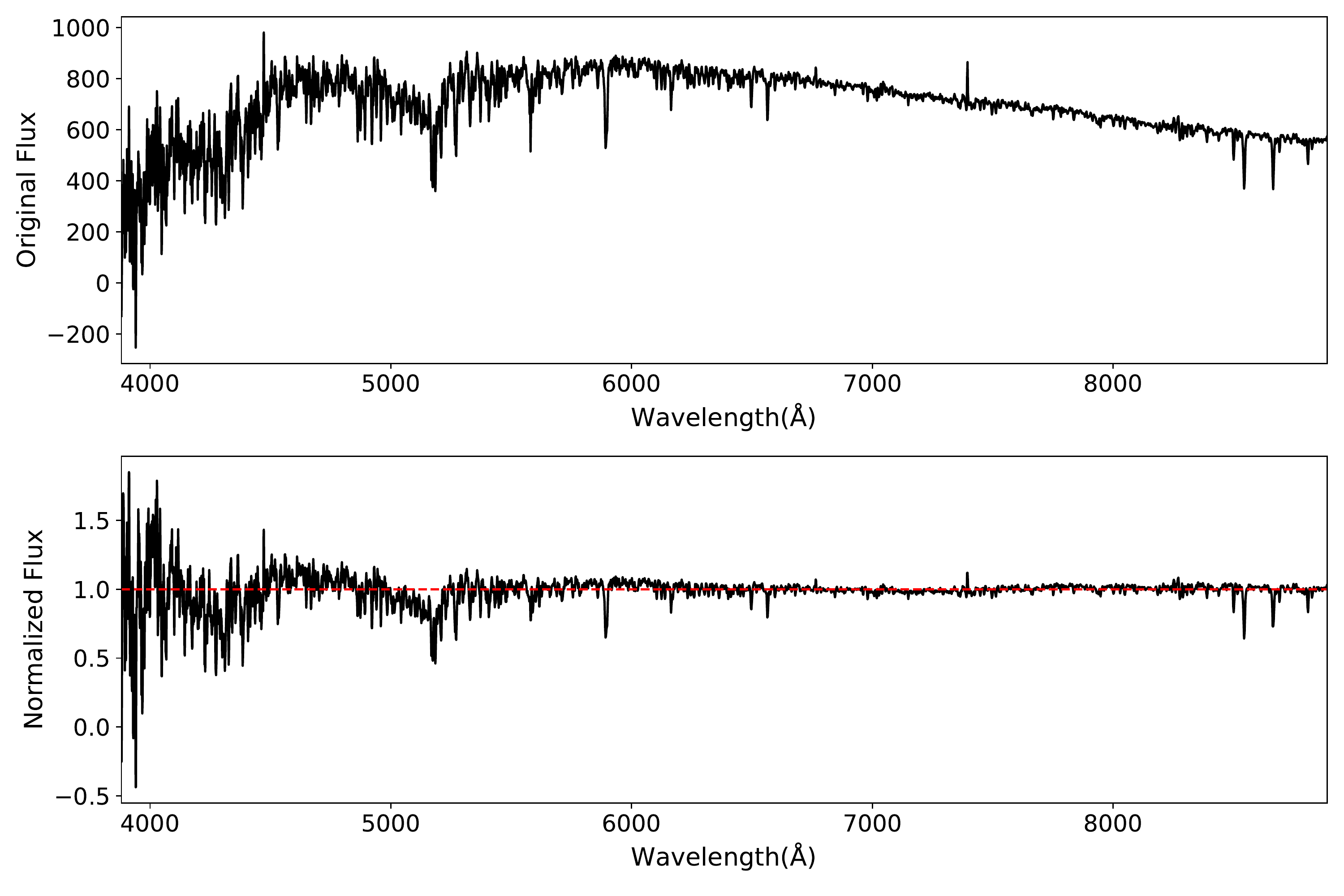}

    \caption{A LAMOST observation (spec-56213-EG000023N024031B01\_sp05-024) and its normalization result. The upper panel and the lower panel respectively present the original LAMOST spectrum, and the continuum-normalized spectrum. The proposed parameter estimation model is experimented on the continuum-normalized spectrum in the bottom panel.}
    \label{fig:normalization}
\end{figure*}

After the above procedures, we obtained the normalized spectra, and the subsequent feature selection and parameter estimation are all conducted on the normalized spectra. An example of continuum normalization is presented in Figure~\ref{fig:normalization}.

\section{THE LASSO-MLPNet}
\label{sec:method}

Machine learning algorithms can automatically learn a specific rule from reference data and this learnt rule can be used to predict new data. In particular, the Neural Network (NN) is a typical machine learning algorithm which has received widely attentions in spectral parameter estimation. The NN can automatically obtain an approximation of the mapping from spectral features to the atmospheric parameters by learning the given reference data. In case of inputting a stellar spectra into the learned NN model, an estimation of an atmospheric parameter can be computed.

This work proposed a Multilayer Perceptron neural network (MLPNet) based on LASSO features for estimating stellar atmospheric parameters. The scheme firstly uses the LASSO algorithm to adaptively evaluate the effectiveness of the fluxes in estimating atmospheric parameter in the existence of high-level noises, and selects the parameter-sensitive spectral features accordingly. Based on the selected spectral features, a Multilayer Perceptron neural network is used to estimate the atmospheric parameters.

\subsection{LASSO}

LASSO \citep{tibshirani1996regression} is a constrained biased estimator, used for feature selection. The purpose of LASSO feature selection is to improve the accuracy and interpretability of subsequent parameter estimation \citep{li2014sdss,li2015linearly}. The LASSO can detect the spectral features sensitive to stellar atmospheric parameters, and eliminate the redundant and invalid data components by constructing a penalty function. This penalty function fuses the objectives of the parameter estimation accuracy and the spareness of selected features. Supposing that we have a reference set $\{(\boldsymbol{x}_i, y_i), i = 1, \cdots, N\}$, where $x_i=(x_{i1},x_{i2},\dots,x_{in})^{\mathrm{T}}$ is the $i$-th input variable (a spectrum), and $y_i$ is the output parameter (the estimated atmospheric parameter). We assume that $x_{i j}$ are standardized, where $\sum_{i}x_{i j}=0, \sum_{i}x_{i j}^2/N = 1$. The penalty function of the LASSO algorithm is:

\begin{equation}
  \begin{aligned}
&(\hat{b}, \hat{\beta})=\mathop{\arg\min}_{(b, \beta)}\sum_{i =1}^{N}\Big(y_i - b - \sum_{j}\beta_{j} x_{i j}\Big)^{2},\\
&s.t. \sum_{j}|\beta_{j}| \leq t,
	\label{eq:lasso}
  \end{aligned}
\end{equation}
where $\beta_{j}$ is the coefficient of the input variable and $t$ is a non-negative parameter that controls the sparsity of the model. Due to the differences on the correlations between these three atmospheric physical parameters and stellar spectra, this work extracted the spectral features using the LASSO independently for $T_{\mathrm{eff}}$, $log~g$ and [Fe/H].

\subsection{Neural Network}

The estimation of the stellar atmospheric parameters based on the Artificial Neural Network (ANN) method has been investigated in the pioneering work \citep{bailer1997physical}. ANN is a computational model (Figure ~\ref{fig:Neural_Network}) composed of a series of hierarchically organized computational units, and each computational unit is referred to as a neuron in related literatures. This work proposed a Multilayer Perceptron neural network (MLPNet) for estimating atmospheric parameters. This MLPNet consists of an input layer, two hidden layers and an output layer. The input to the neural network is the LASSO features of the LAMOST stellar spectrum, and the output is the atmospheric parameter to be estimated. Suppose that $y_i^k$ and $y_{j}^{k+1}$ respectively represents the outputs of the $i$-th node in the $k$-th layer and the $j$-th node in the $(k+1)$-th layer of the neural network, and $w^k_{ij}$ represents the weight on the connection between the above two nodes. If the $(k+1)$-th layer is a hidden layer, the relationship between the node outputs of the two network layers is :

\begin{figure}
\centering
	\includegraphics[width=7cm]{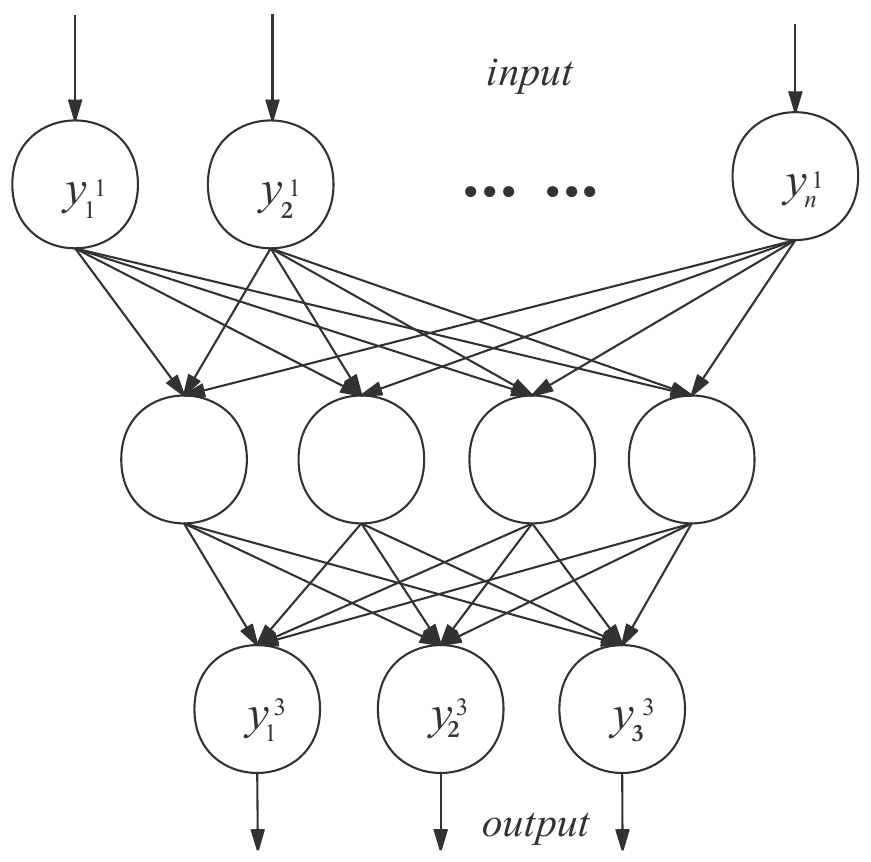}
    \caption{This is a diagram of a neural network, which consists of an input layer, a hidden layer and an output layer.}
    \label{fig:Neural_Network}
\end{figure}

\begin{equation*}
    y_{j}^{k+1}=g\Big(\sum_{i=1}^{n}(w_{i j}^{k}y_{i}^{k} + b_{j}^{k})\Big),
	\label{eq:ann}
\end{equation*}
where $b_j^k$ represents the bias term related with the $j$-th node in the $(k+1)$-th layer, $g$ is an activation function. This work used the sigmoid activation function:
\begin{equation*}
	\sigma(x) = \frac{1}{1 + e^{-x}}.
	\label{eq:Logistic}
\end{equation*}
The activation function is to improve the learning ability of the nonlinear relationship of the network model. If the $(k+1)$-th layer is the output layer of the parameter estimation network, the relationship between the nodes of the above two layers is:
\begin{equation*}
	y_{j}^{k+1}=\sum_{i=1}^{n}(w_{i j}^{k}y_{i}^{k} + b_{j}^{k}).
	\label{eq:ann_output}
\end{equation*}
Suppose $W$ represents the set of the neural network weights, $B$ the set of neural network biases, $\hat{Y}$ and $Y$ represent the parameter predictions and parameter labels of the reference spectra, respectively. The objective function driving the neural network model to learn is:
\begin{equation*}
	(\hat{W}, \hat{B}) = \mathop{\arg\min}_{\boldsymbol{W, B}}\dfrac{1}{2}||\hat{Y} - Y||_{2}^{2} + \dfrac{\alpha}{2}||W||_{2}^{2},
	\label{eq:loss}
\end{equation*}
where $\alpha$ is the regularization term coefficient, which needs to be specified empirically.

In evaluating the performance of the MLPNet model, this work uses the mean absolute error (MAE), the average error ($\mu$) and the dispersion ($\sigma$, standard deviation of the difference):
\begin{equation}
	\mathrm{MAE} = \dfrac{1}{N}\sum_{i=1}^{n}|\hat{y}_{i} - y_{i}|,
	\label{eq:mae}
\end{equation}
\begin{equation}
	\mathrm{\mu} = \dfrac{1}{N}\sum_{i=1}^{n}(\hat{y}_{i} - y_{i}),
	\label{eq:mu}
\end{equation}
\begin{equation}
	\mathrm{\sigma} = \sqrt{\dfrac{1}{N}\sum_{i=1}^{n}(\hat{y}_{i} - y_{i} - \mu)^2},
	\label{eq:sigma}
\end{equation}
where $\hat{y}_{i}$ and $y_i$ are respectively the model prediction and the reference labels.

\subsection{Training the model LASSO-MLPNet}

In training the LASSO-MLPNet, we used the spectral data from the common stars between LAMOST and APOGEE (more in section \ref{sec:data}). Each reference sample consists of a spectrum from LAMOST DR8 and the corresponding labels from APOGEE\_$payne$ catalog. Because the LASSO and neural network are very sensitive to the scales of spectral features, it is necessary to normalize the spectrum by projecting each flux into the range from 0 to 1 before feature selection.

In the observed fluxes of the LAMOST low-SNR and Low-resolution spectra, there are a series of noises and distortions. And the dimension of the spectra is up to 3600. These factors can negatively affects the computational efficiency and estimation accuracy. Therefore, this work selected the effective spectral features, rejected the redundant and ineffective data components by evaluating the usefulness of the fluxes in the existence of the above-mentioned factors using the LASSO method. The regularization coefficient $t$ in the LASSO controls the sparsity of the model, this paper used a 10-fold cross-validation method to find its optimal configuration, and this operation is implemented in the LASSO package \citep{pedregosa2011scikit}. Table~\ref{tab:dimension} presents the numbers of the selected features, the optimal regularization coefficients for $T_\texttt{eff}$, log~g, and [Fe/H].

\begin{table}
	\centering
	\caption{The numbers of the selected features, the optimal regularization coefficients for $T_\texttt{eff}$, log~g, and [Fe/H] in LASSO.}
	\label{tab:dimension}
	\begin{tabular}{lccr} 
		\hline
		Parameter & $t$  & $d$  \\
		\hline
		$T_{\mathrm{eff}}$ & 0.0002 & 643 \\
		log $g$ & 0.0034 & 775\\

		[Fe/H] & 0.0013 & 930\\
		\hline
	\end{tabular}
\end{table}

Then, we trained a MLPNet using the extracted spectral features. The model parameters of the Neural Network, such as the weights and biases, are learnt using a back propagation algorithm. In this work, a four-layer network model was separately trained for each stellar atmospheric parameter. These four layers are an input layer, two hidden layers and an output layer. This network took the stellar spectral features extracted by LASSO as input, and the output layer has only one node, which represents the parameter to be estimated. During training the model, we conducted 2,000 iterations and adopted an early stop method to prevent the model from overfitting.

In conclusion, the training of LASSO, MLPNet, and the determination of the hyper-parameters in LASSO, MLPNet and preprocessing procedures are based on the training set (sec. \ref{sec:data}).

\section{EXPERIMENTS}
\label{sec:experiment}

\subsection{Consistencies with the APOGEE\_$payne$ catalog}\label{sec:experiment:Apogee}

\begin{figure*}
	\includegraphics[width=0.8\textwidth]{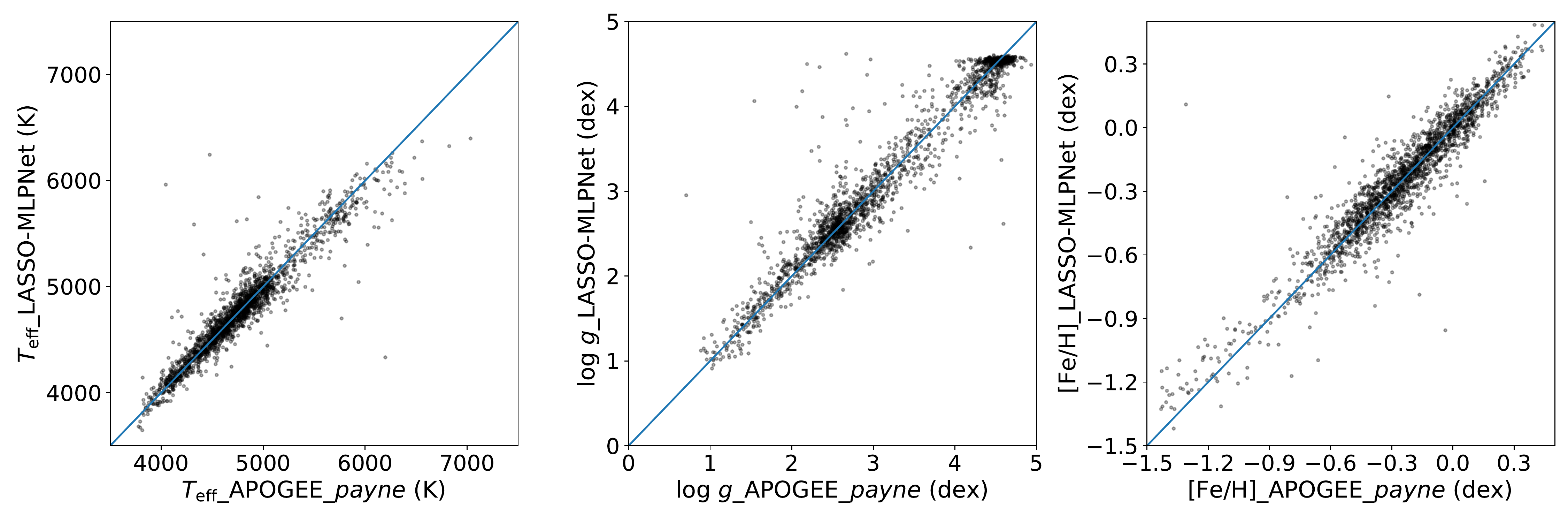}
	\includegraphics[width=0.8\textwidth]{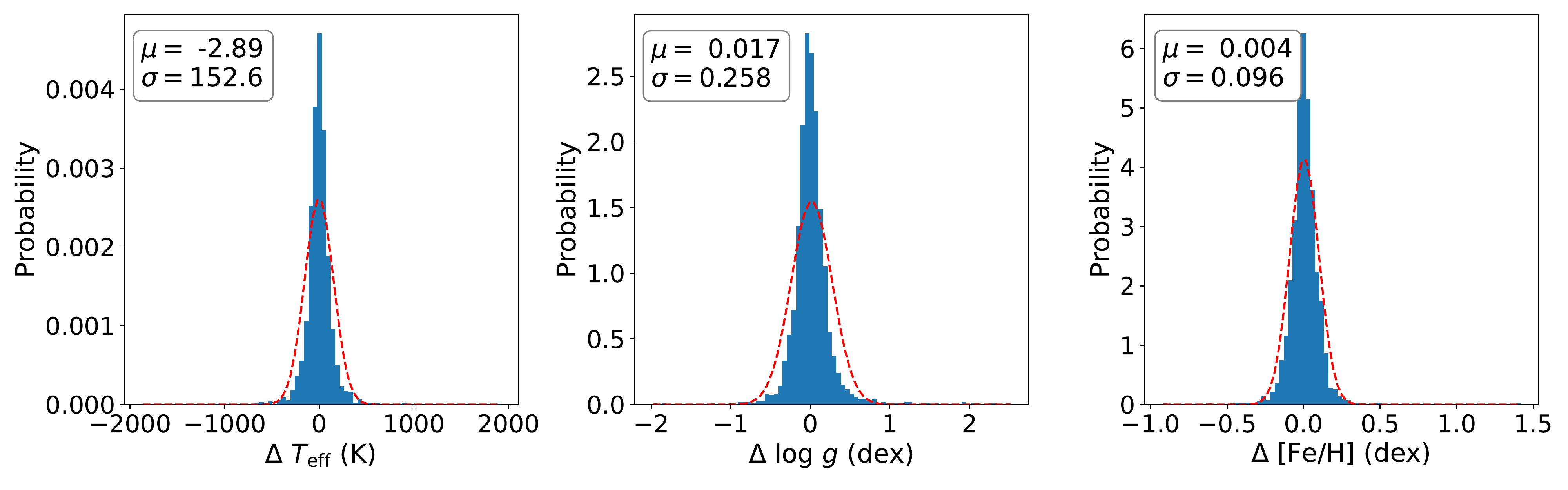}

    \caption{Consistencies between the prediction from MLPNet and the APOGEE\_$payne$ catalog. This figure presents the results on 1,975 test spectra (test set 1) of common objects between LAMOST and APOGEE. The upper panel is the scatter diagram
  with the theoretical consistency in solid line. The lower panels present distribution of residual error on the test set 1. The red dashed curve is a Gaussian fitting.}
    \label{fig:MLPNet_APOGEE_payne}
\end{figure*}

The LASSO-MLPNet model is learnt from the training set (sec. \ref{sec:data}) for predicting ($T_{\mathrm{eff}}$, log $g$ and [Fe/H]) from LAMOST spectra. The consistencies of the LASSO-MLPNet predictions and the APOGEE\_$payne$ catalog can be investigated using a scatter diagram on the test set (Figure~\ref{fig:MLPNet_APOGEE_payne}). In Figure~\ref{fig:MLPNet_APOGEE_payne}, the upper panel presents the scatter diagram for stellar atmospheric parameters $T_{\mathrm{eff}}$, log $g$ and [Fe/H]. The lower panel presents the histogram of the difference between the LASSO-MLPNet estimations and APOGEE\_$payne$ catalog for each parameter. It is shown that the scatter points of three parameters are near the theoretical consistency line. Therefore, there exists excellent consistency between LASSO-MLPNet estimation and the APOGEE\_$payne$ catalog.

\begin{table}
	\centering
	\caption{The consistencies between the LASSO-MLPNet predictions and the APOGEE\_$payne$ catalog. These experimental results are computed from the test set 1. }
	\label{tab:MAE_MLPNet_APO}
	\begin{tabular}{lccr} 
		\hline
		Parameter & MAE & $\mu$ & $\sigma$\\
		\hline
		$T_{\mathrm{eff}}$ (K) & 90.29 & -2.892 & 152.6\\
		log $g$ (dex) & 0.152 & 0.0171 & 0.258\\
		
		[Fe/H] (dex) & 0.064 & 0.0044 & 0.096\\
		\hline
	\end{tabular}
\end{table}

The consistencies between the LASSO-MLPNet prediction and the APOGEE\_$payne$ catalog can also be reflected by the statistics (MAE, $\mu$ and $\sigma$ (equations \ref{eq:mae}, \ref{eq:mu},\ref{eq:sigma}) on the differences between LASSO-MLPNet prediction and APOGEE\_$payne$ catalog (Table~\ref{tab:MAE_MLPNet_APO}). The $\mu$ is the mean of the differences between LASSO-MLPNet prediction and APOGEE\_$payne$ catalog, reflects the systematic offset/consistency between the them; The MAE is the cumulative measurement of the difference on the test sample, this statistical variable depicts the overall inconsistency; The $\sigma$ measures the mean of square error of the difference between the predictions from LASSO-MLPNet and the APOGEE\_$payne$ catalog, and measures the stability of this consistency. In Table~\ref{tab:MAE_MLPNet_APO}, the results show that the LASSO-MLPNet model performs well in terms of systematic bias, overall consistency and stability. But it is worth noting that $T_{\mathrm{eff}}$ is mainly distributed on [4000, 6500] K and [Fe/H] is distributed on [-1.2, 0.439] dex, and APOGEE lacks spectral data for cold stars, hot stars and low-metal stars. These factors may affect the generalization ability of the model in these cases. Therefore, we should be careful when using the results from the LASSO-MLPNet in the cases out of the above-mentioned ranges.

\begin{figure*}
	\includegraphics[width=0.49\textwidth]{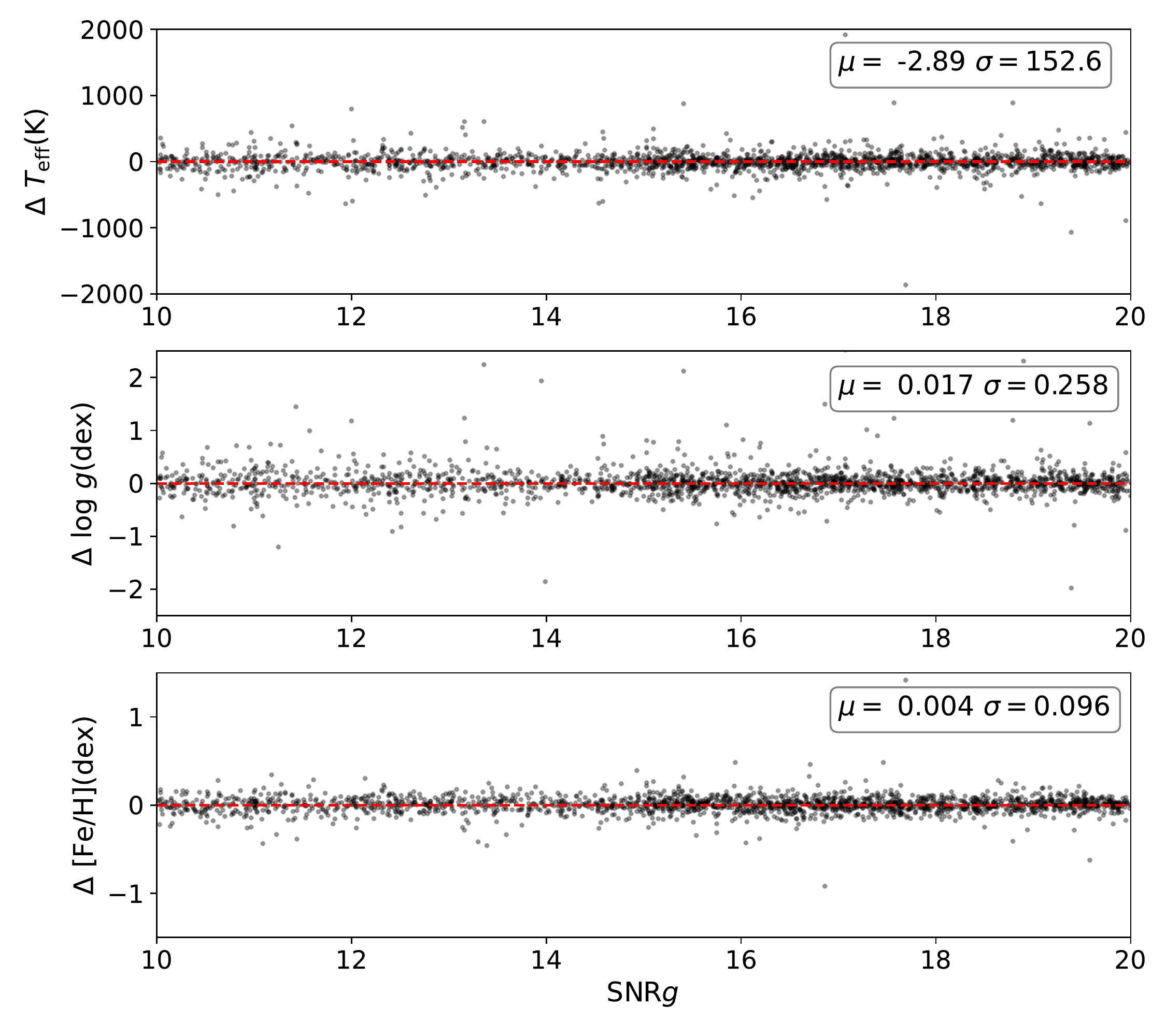}
	\includegraphics[width=0.49\textwidth]{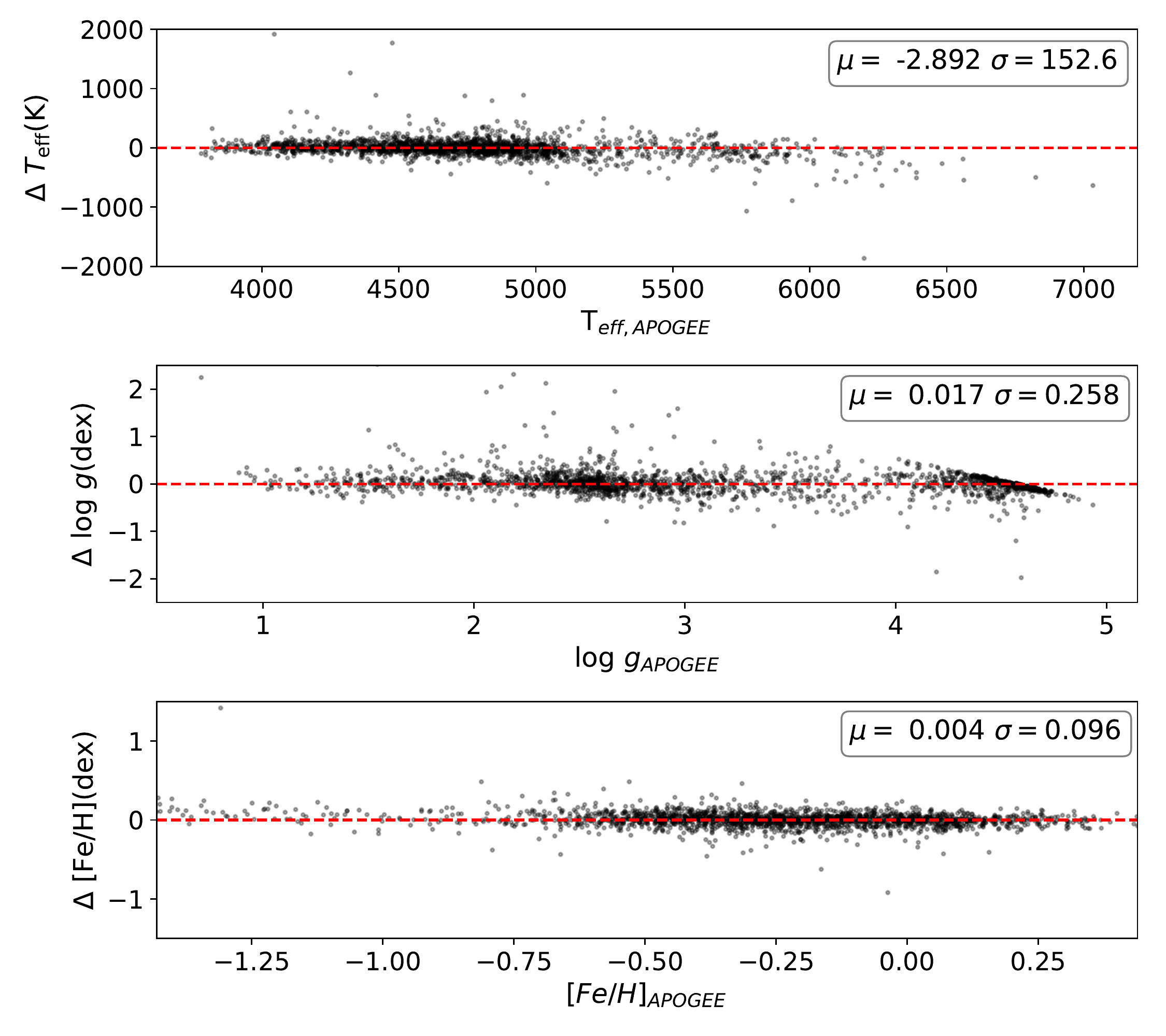}	
	\caption{The dependencies of the consistency between LASSO-MLPNet predictions and the APOGEE\_$payne$ catalog on SNR$g$ and the atmospheric  parameters. These experimental results are computed from the test set 1.The left panel presents the dependencies of the consistency on SNR$g$. The right panel presents the dependencies of the consistencies on $T_\texttt{eff}$, $\log ~g$, and [Fe/H]. The mean and standard deviation of the differences are shown in the figure.}
	\label{fig:MLPNet_APOGEE_payne_diff}
\end{figure*}

The consistencies between the MLPNet predictions and the APOGEE\_$payne$ catalog can also be studied by exploring the dependence of difference between the two results on the SNR and the parameters to be estimated (Figure~\ref{fig:MLPNet_APOGEE_payne_diff}). The left panel shows the differences as a function of SNR$g$. It does not show any visible correlation between the differences and the SNR$g$. On the whole, the difference between the LASSO-MLPNet prediction and the APOGEE\_$payne$ catalog are around 0, which indicates that there is strong consistency between them. The right panel shows the differences as a function of the stellar atmospheric parameter. For $T_\texttt{eff}$, when $T_\texttt{eff}$ < 6000 K, the differences are uniformly distributed near 0, and when $T_\texttt{eff}$ > 6000 K, the distribution of differences tends to shift downward, which means that $T_\texttt{eff}$ derived from stellar spectra is underestimated. This situation results from the scarceness of spectral data in this range. This scarceness makes the model difficult to be sufficiently trained. For log$g$ and [Fe/H], although there are several samples with relatively large differences, the whole is uniformly distributed near 0. In summary, these experimental results show that there are good agreement between the LASSO-MLPNet prediction and APOGEE\_$payne$ catalog.

\subsection{Comparisons with other typical regression methods}

\begin{table*}
	\centering
	\caption{Comparing the LASSO-MLPNet with other typical regression methods. LinearR: Linear Regression, ElasticNet, SVR(rbf): a Support Vector Machine with a rbf kernel, W-SVR(rbf): Weighted SVR(rbf), RandomForest: Random Forest Regressor, ExtraTrees: Extreme Tree Regressor, MLPNet: the method proposed in this work. }
	\label{tab:different_methods}
	\begin{tabular}{lccccccccr} 
		\hline
		\multirow{2}{*}{method}&  & $T_{\mathrm{eff}}$ &  &  & log $g$ & &  & [Fe/H] &  \\\specialrule{0em}{1.5pt}{1.5pt}
		\cline{2-10}
		\specialrule{0em}{1.5pt}{1.5pt}
		& MAE & $\mu$ & $\sigma$ & MAE & $\mu$ & $\sigma$ & MAE & $\mu$ & $\sigma$\\
		\hline
		LinearR & 113.55 & -0.471 & 202.76 & 0.270 & -0.055 & 2.960 & 0.099 & 0.0051 & 0.377\\
		ElasticNet & 109.73 & -1.970 & 190.68 & 0.254 & -0.0431 & 2.424 & 0.096 & -0.0051 & 0.388\\
		SVR(rbf) & 375.23 & 198.71 & 414.44 & 0.228 & 0.0070 & 0.356 & 0.080 & -0.0046 & 0.118\\
		W-SVR(rbf) & 323.73 & 195.86 & 343.18 & 0.175 & -0.0041 & 0.269 & 0.075 & -0.0036 & 0.110\\
		RandomForest & 107.71 & -2.083 & 170.41 & 0.207 & -0.0051 & 0.317 & 0.105 & -0.0054 & 0.143\\
		ExtraTrees & 102.60 & -2.112 & 163.67 & 0.191 & -0.0070 & 0.287 & 0.099 & -0.0059 & 0.137\\
		\textbf{MLPNet} & \textbf{90.29} & \textbf{-2.892} & \textbf{152.65} & \textbf{0.152} & \textbf{0.0171} & \textbf{0.258} & \textbf{0.064} & \textbf{0.0044} & \textbf{0.096}\\
		\hline
	\end{tabular}
\end{table*}

To evaluate the proposed scheme, this work also compared the LASSO-MLPNet with the following typical regression methods: Linear Regression (LinearR), ElasticNet, Support Vector Machine Regression (SVR), Random Forest Regression (RandomForest) and Extreme Trees Regression (ExtraTrees), etc. In these comparisons, we replaced the MLPNet with LinearR, ElasticNet, SVR, RandomForest and ExtraTrees, and keep other configurations the same as the experiments in section \ref{sec:experiment:Apogee}. For example, training set and test set 1 , spectral preprocessing procedures and LASSO features and so on (Table~\ref{tab:dimension}). The optimization of the parameters is implemented through a grid search, and the experimental results are presented in Table ~\ref{tab:different_methods}.

The performance of the above-mentioned methods in parameter estimation can be investigated using the difference between their predictions and the APOGEE\_$payne$ catalog. Table~\ref{tab:different_methods} presents the estimation difference of the six methods on test set 1. On the whole, the MLPNet method outperforms the other methods. For linear methods, the estimation performance of ElasticNet is better than LinearR. This is because there are certain noises and calibration defects in the observed spectra, and LinearR without regularization is more sensitive to such noises and defects. Moreover, nonlinear methods (ExtraTrees, RandomForest, MLPNet) perform better than linear methods (LinearR, ElasticNet) in estimating stellar atmospheric parameters. These results indicate that there exists some non-linear relationships between the spectral features and three stellar atmospheric parameters ($T_{\mathrm{eff}}$, log $g$ and [Fe/H]). Therefore, it is more suitable to derive the atmospheric parameters using a non-linear regression method. The ExtraTrees and RandomForest are highly parallel algorithms. Among them, RandomForest learns by randomly sampling the training set and selects the best splitting properties from a random subset, while the ExtraTrees uses the entire training set and randomly selects a splitting subset of properties to construct a decision tree. These characteristics make the ExtraTrees model possible to have stronger randomness, smaller model variance ($\sigma^2$), and better generalization performance. However, the offset ($\mu$) of ExtraTrees may increase (Table~\ref{tab:different_methods}). For [Fe/H], the errors from ExtraTrees and RandomForest are higher than those of the linear model. The possible reason is that the spectra with low SNR contains more noise, while their tolerance for noise is relatively low, which can result in overfitting.

The estimation performance of SVR with a rbf kernel is better than linear models (LinearR, ElasticNet) on log $g$ and [Fe/H]. However, on $T_{\mathrm{eff}}$ estimation, the error from SVR(rbf) is too large. It is known that the SVR method treats each input feature equally, but they may be different from each other on their contributions to the parameter estimation in reality. This inappropriate treatment may cause masking effects on features. Therefore, we investigated a reweighting scheme, W-SVR(rbf), on spectral features. The feature weights are estimated by LASSO method in selecting features. The W-SVR(rbf) scheme is to train and test the parameter estimation model by inputting the reweighted features into SVR(rbf). The experimental results show that the estimation performance is greatly improved after weighting the spectral features. Furthermore, the performance of W-SVR(rbf) is better than that of ExtraTr and RandomF for log $g$ and [Fe/H] (Table ~\ref{tab:different_methods}). However, for $T_{\mathrm{eff}}$, the estimation error of W-SVR(rbf) is still large. The possible reason is that although the compactness of the spectral features extracted by LASSO is very good, there exists some redundancies and insignificant data components in the selected spectral features for estimating $T_{\mathrm{ eff}}$. These factors lead to a large variance in the estimation results of the model.

In conclusion, the estimation error of the MLPNet model for each atmospheric parameter is smaller than those of other six models. This indicates that the MLPNet has good parameter estimation performance and good predictive capability for low-SNR stellar spectra from LAMOST DR8.

\subsection{Comparison with LASP}
\label{sec:LASP_APOGEE}

To evaluate the accuracy and stability of the LASSO-MLPNet, we computed and analyzed the consistency of LASSO-MLPNet predictions and LASP predictions (LAMOST official catalog) to the APOGEE\_$payne$ catalog. This evaluation was conducted on 1,975 test spectra (test set 1).  LASP\citep{luo2015} implements parameter estimation based on the ULySS method and a stepwise refinement strategy \citep{wu2011automatic}. Actually, the LASP first classifies stars into late A-type and FGK-type stars, then uses the CFI method to obtain initial parameters estimation, and finally generates final atmospheric parameter estimation using ULySS.

\begin{figure*}
	\includegraphics[width=0.8\textwidth]{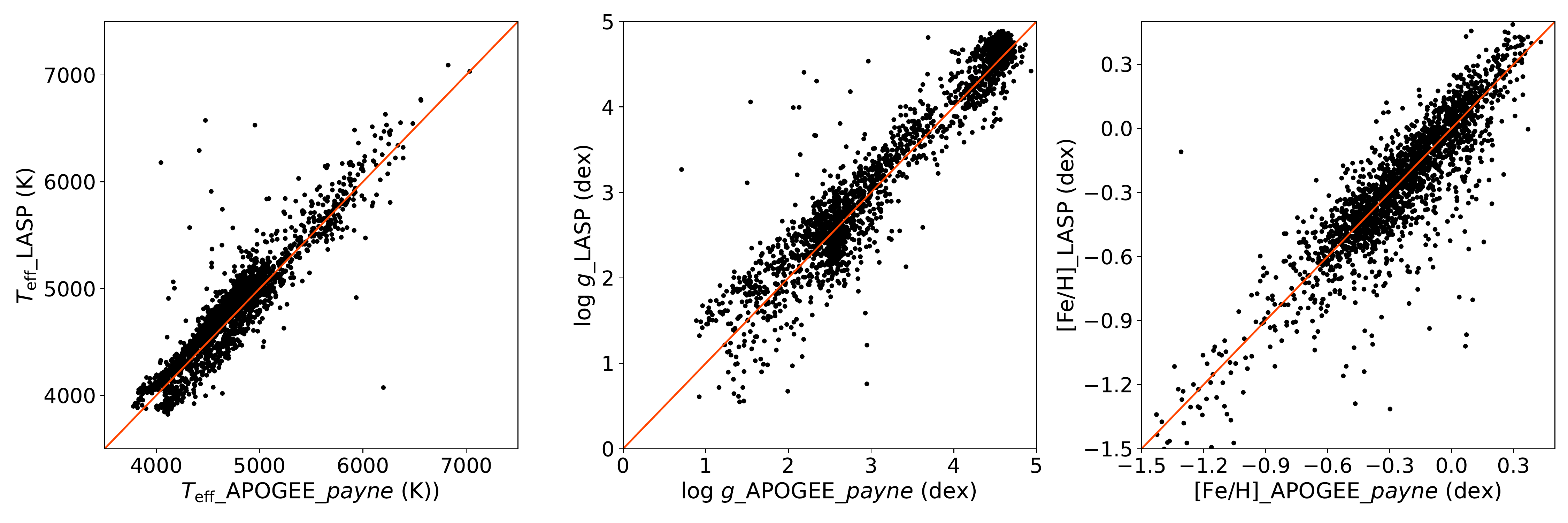}
	\includegraphics[width=0.8\textwidth]{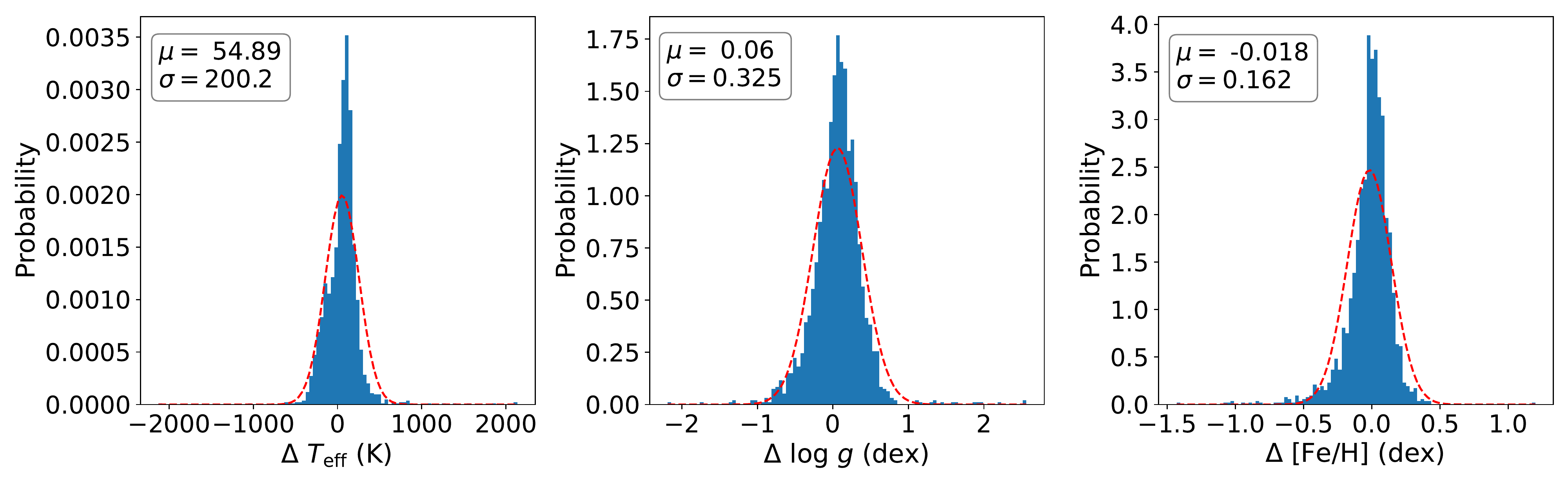}
    \caption{Consistencies between LASP predictions and the AOPGEE\_$payne$ catalog. The upper three subplots are the scatter diagrams with the theoretical consistency in solid line. The lower three subplots present the difference between LASP predictions and the AOPGEE\_$payne$ catalog. The red dashed curve is a Gaussian fitting curve. The evaluation results are calculated on test set 1.}
    \label{fig:LAMOST_APOGEE_payne}
\end{figure*}

The consistencies between the LASP estimations and APOGEE\_$payne$ catalog are investigated using a scatter diagram and a distribution of the difference between the two results (Figure~\ref{fig:LAMOST_APOGEE_payne}). The upper subplots in Figure~\ref{fig:LAMOST_APOGEE_payne} show the scatter plots for 1,975 stellar spectra (test set 1). By comparing the experimental results in Figure~\ref{fig:LAMOST_APOGEE_payne} and Figure~\ref{fig:MLPNet_APOGEE_payne}, it is found that there are more observations with evident inconsistencies between LASP estimations and APOGEE\_$payne$ catalog, and there exist obvious systematic deviations in the LASP estimations from the APOGEE\_$payne$ catalog. Therefore, the LASSO-MLPNet predictions are more consistent with the APOGEE\_$payne$ catalog than LASP predictions. The below subplots in Figure~\ref{fig:LAMOST_APOGEE_payne} show the histogram of differences between LASP estimations and APOGEE\_$payne$ catalog.

\begin{table*}
	\centering
	\caption{The consistencies between LASP predictions, LASSO-MLPNet predictions and the APOGEE\_$payne$ catalog. The left side in the table is the results of the consistency between the LASP estimations and the AOPGEE\_$payne$ catalog (LASP, APOGEE\_$payne$), the right side is the consistency results between the LASSO-MLPNet predictions and the AOPGEE\_$payne$ catalog (LASSO-MLPNet, APOGEE\_$payne$).}
	\label{tab:LASP_APOGEE}
	\begin{tabular}{ccccccc} 
		\hline
		 \multirow{2.5}{*}{parameters}& \multicolumn{3}{c}{(LASP, APOGEE\_$payne$)} & \multicolumn{3}{c}{(LASSO-MLPNet, APOGEE\_$payne$)}  \\
		 \specialrule{0em}{1.5pt}{1.5pt}
		\cline{2-7}
		\specialrule{0em}{1.5pt}{1.5pt}
		 & MAE & $\mu$ & $\sigma$ & MAE & $\mu$ & $\sigma$ \\
		\hline
		$T_{\mathrm{eff}}$ (K) & 144.59 & 54.89 & 200.2 & 90.29 & -2.892 & 152.6 \\
		log $g$ (dex) & 0.236 & 0.060 & 0.325 & 0.152 & 0.0171 & 0.258\\
		
		[Fe/H] (dex) & 0.108 & -0.018 & 0.162 & 0.064 & 0.0044 & 0.096\\
		\hline
	\end{tabular}
\end{table*}

The inconsistencies between LASP estimations, the LASSO-MLPNet predictions and the APOGEE\_$payne$ catalog can also be evaluated by statistical measures MAE, $\mu$ and $\sigma$ (equations \ref{eq:mae}, \ref{eq:mu}, \ref{eq:sigma}, and Table~\ref{tab:LASP_APOGEE}). For each atmospheric parameter, the mean difference $\mu$ and dispersion $\sigma$ from LASP are much evident than the corresponding statistical results from LASSO-MLPNet. Compared with APOGEE\_$payne$ catalog, LASP has an overestimation of 54.89 K for $T_{\mathrm{eff}}$, 0.060 dex for log $g$, and an underestimation of 0.018 dex for [Fe/H]. While, the mean differences of the LASSO-MLPNet, (-2.892 K for $T_{\mathrm{eff}}$, 0.0171 dex for log $g$, and 0.0044 dex for [Fe/H]) are significantly than LASP. Moreover, the dispersion $\sigma$ and the MAE for the LASSO-MLPNet predictions are also much smaller than those from LASP. Therefore, the LASSO-MLPNet model has good parameter estimation capability on the LAMOST low-SNR stellar spectra.

\begin{figure*}
	\includegraphics[width=0.85\textwidth]{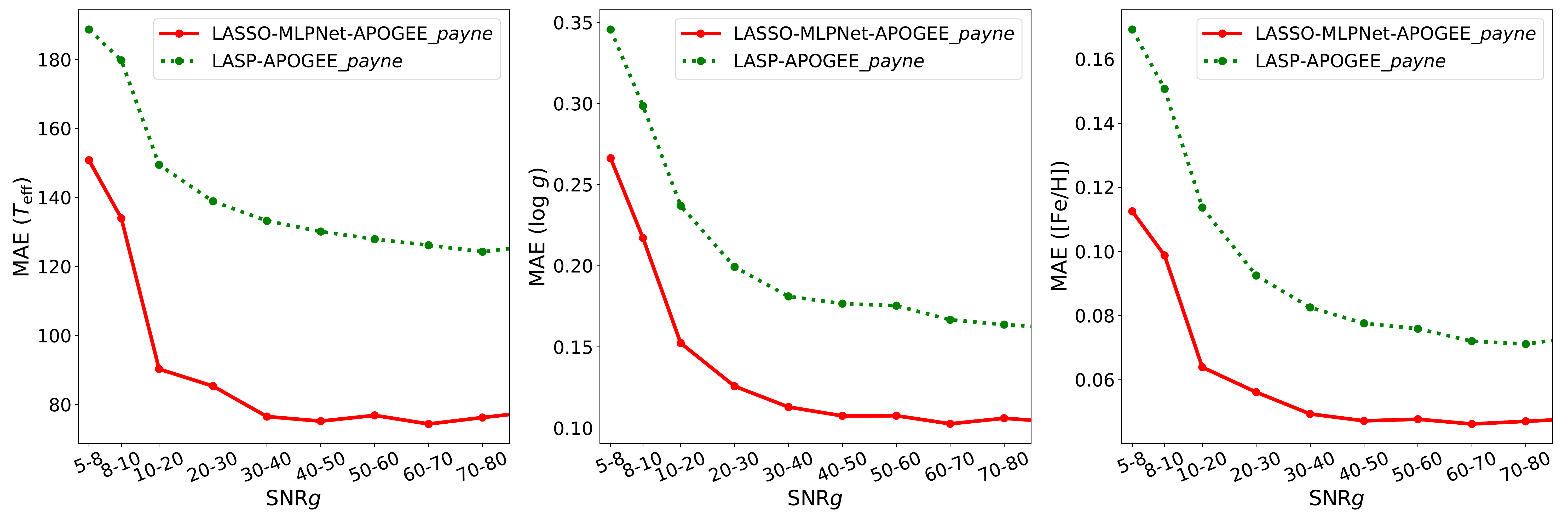}	
	\caption{The consistencies between LASP estimations, LASSO-MLPNet predictions and the APOGEE\_$payne$ catalog. The red solid line and the green dashed line are dependencies of the MAEs between AOPGEE\_$payne$ catalog and LASSO-MLPNet predictions, and the MAEs between the AOPGEE\_$payne$ catalog and LASP estimations, on SNR$g$. The evaluation results of this experiment are calculated on test set 1 and test set 2.}
	\label{fig:LAMOST_APOGEE_payne_SNR}
\end{figure*}

To evaluate the applicability of the LASSO-MLPNet model on a wider range of spectra, we constructed the second test set (test set 2) for the cases 5 < SNR$g$ $\leq 10$ and 20 < SNR$g$ $\leq 80$. This test set contains 56,198 LAMOST stellar spectra. We evaluated the consistency between the predictions from the aforementioned LASSO-MLPNet and the APOGEE\_$payne$ catalog on test set 1 and test set 2 (Figure~\ref{fig:LAMOST_APOGEE_payne_SNR}). The results show that the LASSO-MLPNet improves the accuracy significantly in the case of 5 $<$ SNR $\leq$ 80.

\subsection{Uncertainty analysis}

The uncertainty of the atmospheric parameter estimation is the instability of the estimates caused by observational noise, instrumental effects and parameter estimation model effects. This work investigated the uncertainty from the following two aspects: the integrated uncertainty and the model uncertainty. These uncertainty evaluation results are calculated on test set 1 and test set 2.

\begin{figure*}
	\includegraphics[width=0.85\textwidth]{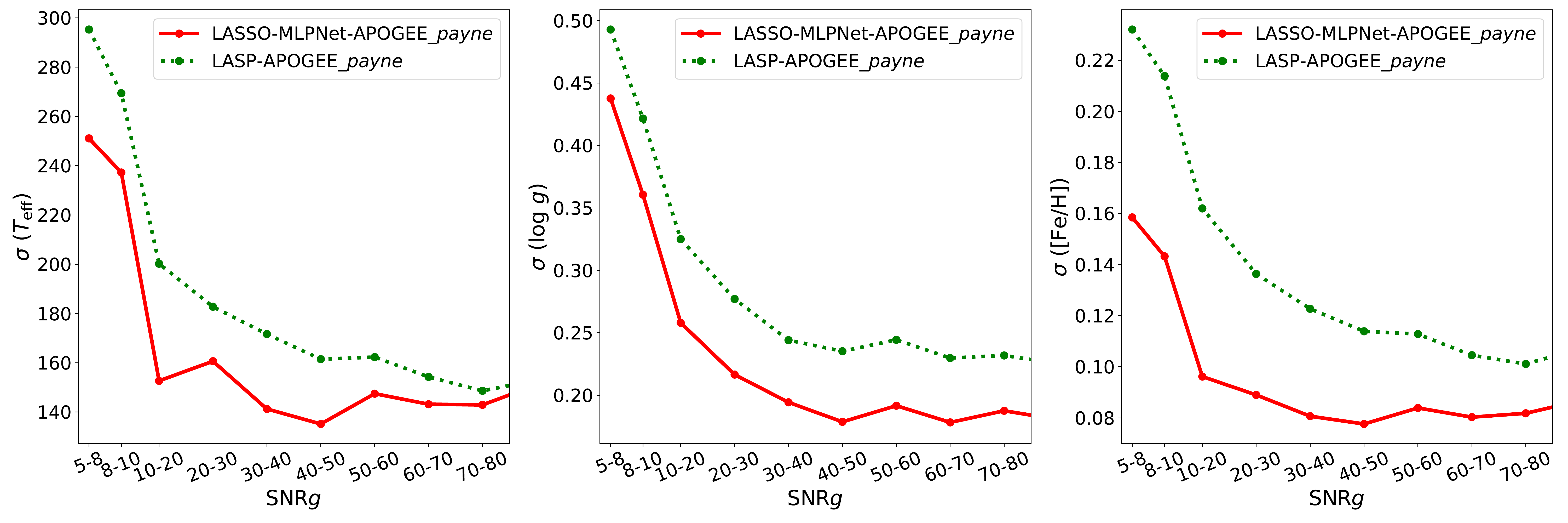}
	\caption{The dispersion $\sigma$ of the difference between the APOGEE\_$payne$ catalog and LASP estimations or the LASSO-MLPNet predictions. The red solid line and the green dashed line are respectively the dependencies of the dispersions of the differences between the APOGEE\_$payne$ catalog and the LASSO-MLPNet predictions, and between the APOGEE\_$payne$ catalog and LASP estimations, on SNR$g$. The evaluation results of this experiment are calculated on test set 1 and test set 2.}
	\label{fig:LAMOST_match_sigma}
\end{figure*}

The integrated uncertainty is measured by the standard deviation of the differences between the model estimations and the reference label. The standard deviation of the difference describes the robustness of the parameter estimation system to some factors, such as noises, instrumental effects and parameter estimation model effects and so on. Figure~\ref{fig:LAMOST_match_sigma} shows the dependencies of the standard deviation/dispersion $\sigma$ of the stellar atmospheric parameters on SNR$g$. The experimental results show that, for the three atmospheric parameters, the dispersions $\sigma$ from the LASSO-MLPNet predictions with APOGEE\_$payne$ catalog are smaller than those from LASP estimations on a wide SNR range. Especially, in case of SNR$g$ $\leq$ 40, the dispersion from the LASSO-MLPNet decreases sharply with the increase of SNR$g$; In case of SNR$g$ > 40, the trend of dispersion tends to be flat. On the contrary, the dispersion of LASP decreases on a much wider SNR range. Therefore, the experimental results also show that, the dispersion from LASSO-MLPNet has a weaker correlation with SNR$g$ than those from LASP. This weaker correlation means that the LASSO-MLPNet model is more robust to the above-mentioned factors. In summary, these experimental results indicate that the parameter estimation performance of the LASSO-MLPNet model is more robust and more certain.
	
As for $T_{\mathrm{eff}}$, however, the dispersion of the difference between the LASSO-MLPNet predictions and APOGEE\_$payne$ catalog increases slightly at the high SNR$g$ range. This phenomenon probably is caused by difference on noise level between training spectra and test spectra. The SNR$g$ of the training data ranges from 10 to 20, these training spectra is disturbed by a lot of noise components. As the quality of the test data (test set 2) increases, the difference between the training data and the test data (test set 2) becomes more and more significant on noise level. Therefore, the dispersion of difference slightly increases in case of the test spectra with very high quality (high SNR$g$). This phenomenon also indicates that it is a potential exploration direction to construct an appropriate training set and the corresponding parameter estimation models respectively for the spectra with different SNR range. On the other hand, this work utilized the spectral fluxes on a more wide wavelength range than the LASP. Although the superiority of the wide wavelength range is more spectral information for parameter estimating in theory, there are also some potential risks, for example, inclusion of a lot of invalid data (e.g., 5700-5900\r A). These kind risks also result in an increase on uncertainty.

\begin{figure*}
	\includegraphics[width=0.87\textwidth]{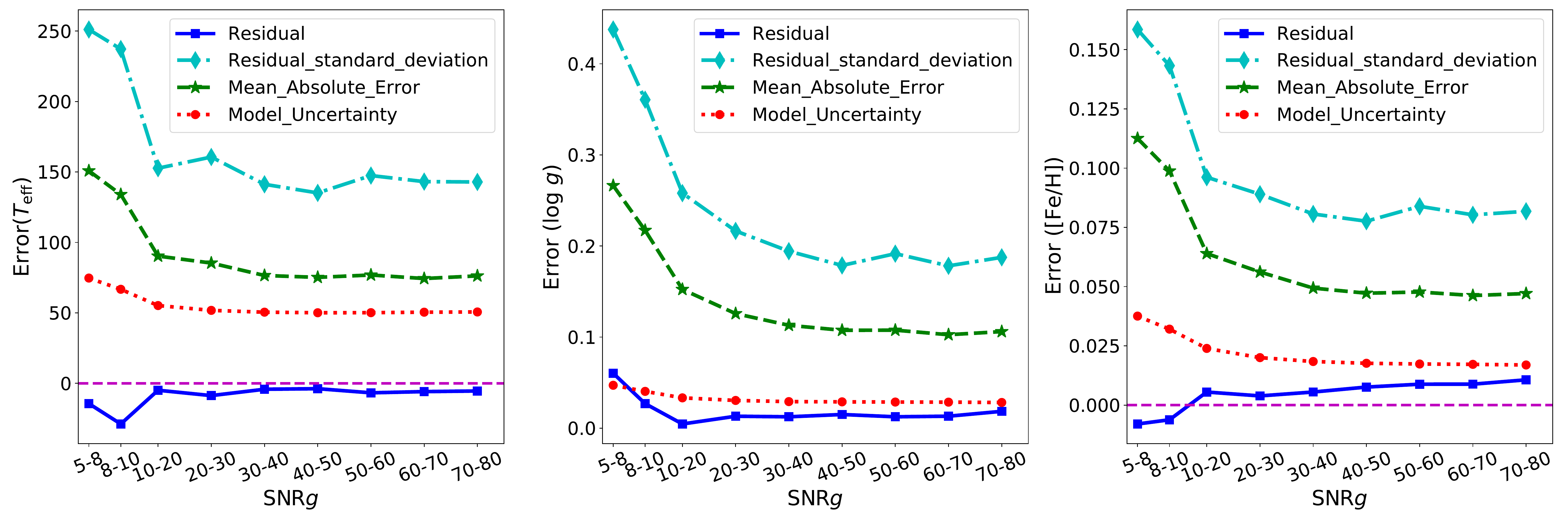}
	\caption{The dependencies of the errors (the residual, the residual standard deviation, the absolute of residual and the model uncertainty) of LASSO-MLPNet predictions on SNR$g$. The residual, the residual standard deviation and the absolute of residual are obtained by calculating the estimations obtained by LASSO-MLPNet method (without dropout). The residual standard deviation is referred to as integrated uncertainty in this work.
	The model uncertainty is obtaind by LASSO-MLPNet method (with dropout). The evaluation results of this experiment are calculated on test set 1 and test set 2.}
	\label{fig:uncertainty-residual-SNR}
\end{figure*}

\begin{figure*}
	\includegraphics[width=0.87\textwidth]{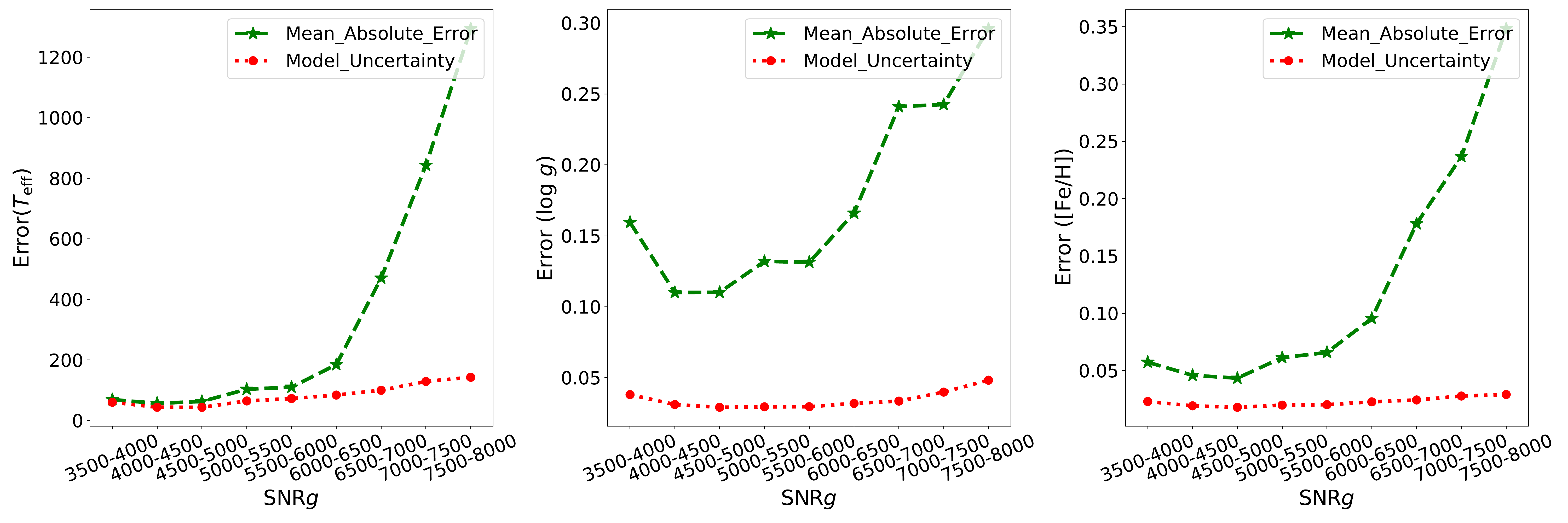}
	\caption{The sensitiveness of the uncertainty to extrapolation. Based on the statistical characteristics of the LAMOST spectra, this work foucs on the parameter estimation of the spectrum with 3500K$\leq T_\texttt{eff}\leq $6500K. It is shown that the model uncertainty is slightly sensitive to the extrapolation. The evaluation results of this experiment are calculated on test set 1 and test set 2.}
	\label{fig:uncertainty-extrapolation}
\end{figure*}

The model uncertainty results from the conditions of the quantity and parameter coverages of the training spectra, the randomness of the obtained model parameters, and so on. This work used the Dropout technique \citep{hron2018variational} to estimate the model uncertainty of LASSO-MLPNet. Related researches show that the Dropout technique is a Bayesian approximation of the model uncertainty \citep{gal2016dropout}. We trained 10 different LASSO-MLPNet models, which can give ten different predictions for each test spectrum. The standard deviation of these 10 predictions is calculated as the model uncertainty measurement (the red dotted line) of the LASSO-MLPNet predictions on that spectrum (Figure~\ref{fig:uncertainty-residual-SNR}). The experimental results show that, the model uncertainty decreases with the increase of SNR$g$.

To further investigate the rationality of the uncertainty measures, this work studies the dependencies of the estimation resiudal,  mean absolute error (absolute residual), residual standard deviation (integrated uncertainty) and model uncertainty on residual (Figure~\ref{fig:uncertainty-residual-SNR}). It is shown that, on the whole there exist some positive correlations between the proposed uncertainty measures (the integrated uncertainty and the model uncertainty) and the estimation inconsistency (residual, mean absolute error/residual). These positive correlations indicate the rationality of the uncertainty measures. However, some of the residuals are positive, and the others are negative. The changing sign of residual brings some difficulties in reading this positive correlation. Therefore, we also plotted the dependencies of the absolute of residual on SNR (Figure~\ref{fig:uncertainty-residual-SNR}). With the increase of SNR, the standard deviation of residual, the absolute of residual and model uncertainty gradually decrease in general. Specifically, they decreases sharply with the increase of SNR in case of SNR$g$ < 20. While in case of SNR$g$ > 20, the downward trend tends to be flat. Therefore, the model uncertainty correctly indicates the reliability of the estimated parameters.

We also investigated the sensitiveness of the proposed uncertainty measures to the extrapolation (Figure~\ref{fig:uncertainty-extrapolation}). Based on the characteristics of the LAMOST observations, this work focused on the parameter estimation of the spectrum with 3500K$\leq T_\texttt{eff}\leq $6500K. It is shown that the model uncertainty is slightly sensitive to the extrapolation. In case of $T_{\mathrm{eff}} \leq 6500$ K, the uncertainty of the model and the absolute of parameter residual have no obvious trend  with the increase of $T_{\mathrm{eff}}$. However in case of $T_{\mathrm{eff}} > 6500$ K, the model uncertainty has an slight upward trend. This indicates that the model uncertainty can alarm the cases of parameter estimation performance degradations.

Some uncertainties come from noises and reference label quality variations. This work studied the evaluation of the uncertainty of the proposed scheme, but not tried to reduce the negative influences from the uncertainties of noises and reference label quality variations. Actually, \cite{leung2019deep} designed a novel objective function to reduce these kinds of negative influeces from label noises and observation noises. This is an interesting and valuable investigation. We will study them in the next step.

\section{Application to LAMOST Stellar Spectra}
\label{sec:application}

In Section~\ref{sec:experiment}, we evaluated the LASSO-MLPNet model on test set 1 and test set 2, and conducted a series of  comparisons and analysis on the performance of this model. It is shown that the proposed model is robust and applicable to a wide range of stellar spectra. Therefore, this work applied the LASSO-MLPNet model to estimating atmospheric parameters from LAMOST low-resolution stellar spectra with $5< \mathrm{SNR}g \leq 80$. In this SNR range, the spectra without the estimations of redshift and atmospheric parameters in the LAMOST DR8 catalog is excluded.

\begin{figure}
	\centering
	\includegraphics[width=8cm]{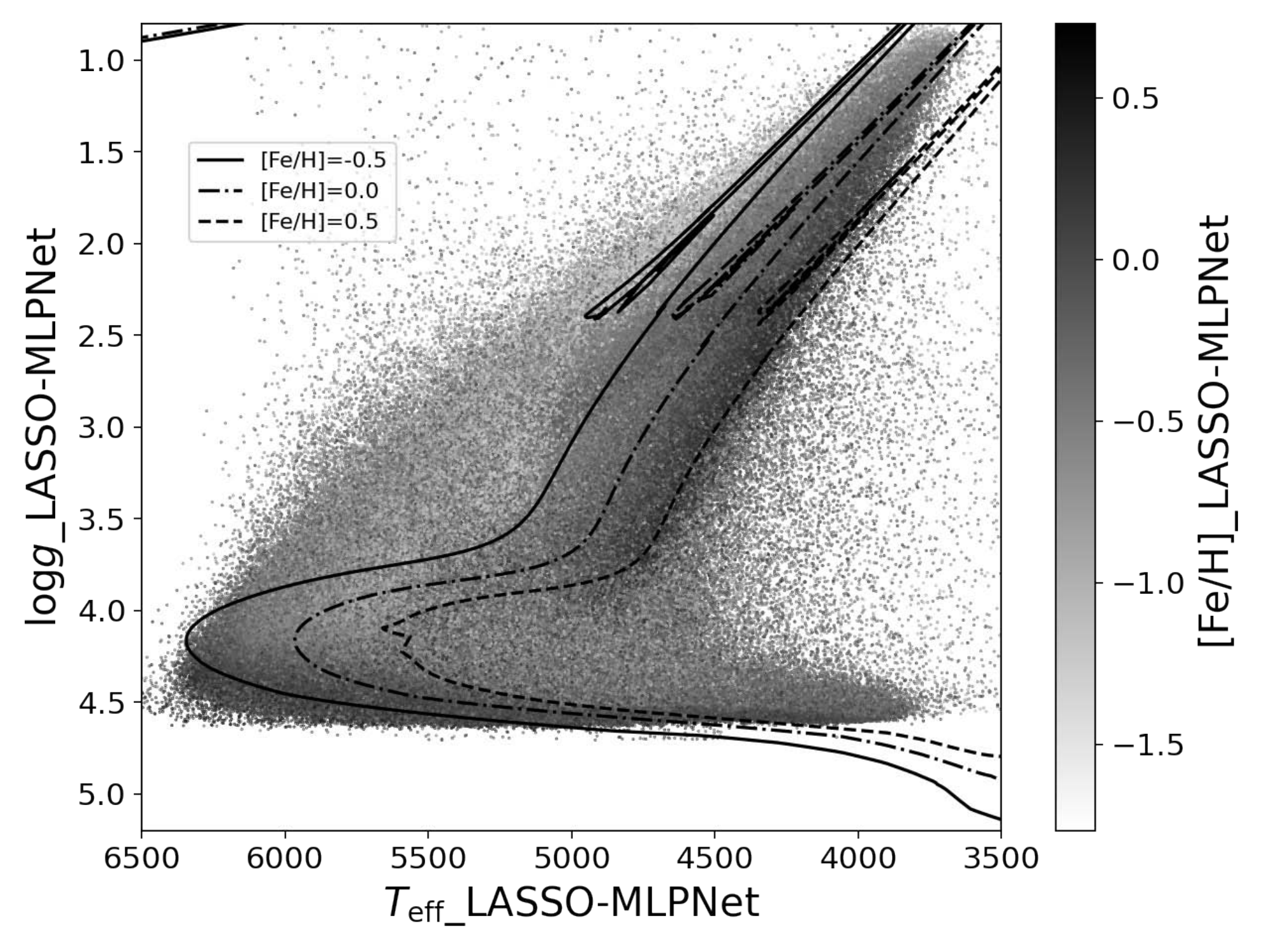}
	\caption{The distribution diagram of the LASSO-MLPNet catalog on $T_{\mathrm{eff}}$-log $g$. The color indicates the value of [Fe/H]. The isochrones are computed from the MIST stellar evolution model, in which the stellar age is 7 Gyr, and the [Fe/H] is -0.5 dex (solid line), 0 (dotted line) and 0.5 dex (dotted line), respectively.}
	\label{fig:MLPNet_482}
\end{figure}

The APOGEE survey mainly contains G and K-type stars, and lacks hot and cold stars. Therefore,  there exist very little observations with $T_{\mathrm{eff}}<3500K$, $T_{\mathrm{eff}}>6500K$ from common stars between LAMOST and APOGEE\_$payne$ catalog. The result is that the reference data on this parameter ranges are insufficient to train a LASSO-MLPNet model. Therefore, the spectra in this parameter range are also excluded from the processing of the LASSO-MLPNet model. Therefore, we estimated the stellar atmospheric parameters for 4,828,190 LAMOST DR8 stellar spectra with 3500 $\mathrm{K}\leq T_{\mathrm{eff}} \leq 6500$ K and $5 < \mathrm{SNR}g \leq 80$. The estimations are released as a value-added catalog. The distribution map of the predictions is presented in Figure~\ref{fig:MLPNet_482}. The isochrones in this figure are the stellar evolution  trace from MIST with a stellar age of 7Gyr \citep{choi2016mesa,dotter2016mesa}.

The proposed LASSO-MLPNet is data-driven method. The learning and performance of it depend on sufficiencies of reference data in training set. Therefore, the LASSO-MLPNet can degrade in the parameter ranges with scarce reference data, for example, the spectra with  $T_{\mathrm{eff}}<3500K$, $T_{\mathrm{eff}}>6500K$, and [Fe/H]$<$ -1.5. Therefore, the released catalog does not contain the information of this kind spectra. On the contrary, the LASP delivers stellar parameters by minimizing the $\chi^2$ between observed spectra and template spectra from ELODIE library \citep{prugniel2001database, prugniel2007new}. This is a kind of prototype methods, which are relatively excellent on the parameter ranges with less reference samples.

\section{Conclusions}
\label{sec:conclusions}

In this article, A Multilayer Perceptron neural network model based on LASSO (LASSO-MLPNet) was proposed for estimating atmospheric parameters ($T_{\mathrm{eff}}$, log $g$ and [Fe/H]) from LAMOST DR8 spectra. We used the stellar spectra from common stars between LAMOST and APOGEE\_$payne$ catalog as the reference data. Each reference sample consists of a LAMOST DR8 spectrum and the corresponding parameter estimations from the AOPGEE\_$payne$ catalog. The experimental results show that, the MLNet predictions are excellently consistent with the APOGEE\_$payne$ catalog. In case of the spectra with $10 < \mathrm{SNR}g \leq 20$, the MAEs of three atmospheric parameters ($T_{\mathrm{eff}}$,  log $g$ and [Fe/H]) are 90.29 K, 0.152 dex, 0.064 dex, respectively.

To evaluate the predictive performance of the proposed model, this paper compared LASSO-MLPNet with six typical regression methods. The experimental results show that our model has good estimation capability. At the same time, we also evaluated the consistencies between the APOGEE\_$payne$ catalog and LASP estimations. The accuracy of the three parameters are 145.4 K, 0.236 dex, 0.110 dex, respectively. Compared with the results in Section~\ref{sec:experiment}, the LASSO-MLPNet predictions shows better consistency with the APOGEE\_$payne$ catalog. And we analyzed the uncertainty of parameter estimation, and the results show that the LASSO-MLPNet model has good robustness.

Due to the lack of cold and hot stars in APOGEE, the available samples of this kind are scarce in the reference data. Therefore, the generalization performance of the model may be relatively inferior on the spectra of these stars. The spectra with low SNR are contaminated with much noise, and the metal lines on them are very weak. These factors directly affect the estimation performance of the LASSO-MLPNet. Therefore, there should be a lot of works under being conducted in future to improve the parameter estimation performance further for these low-quality spectra.

As part of the study, we estimated atmospheric parameters for more than 4.82 million low-resolution spectra with $5<\mathrm{SNR}g\leq80$ and 3500 $\mathrm{K}\leq T_{\mathrm{eff}} \leq 6500$ K from LAMOST DR8. The estimations results are released as a value-added catalog for reference.

\section*{Acknowledgements}

This work were supported by the National Natural Science Foundation of China (Grant No. 11973022), the Natural Science Foundation of Guangdong Province (No. 2020A1515010710), the Major projects of the joint fund of Guangdong and the National Natural Science Foundation (Grant No. U1811464).

LAMOST, a multi-target optical fiber spectroscopic telescope in the large sky area, is a major national engineering project built by the Chinese Academy of Sciences. Funding for the project is provided by the National Development and Reform Commission. LAMOST is operated and managed by the National Astronomical Observatory of the Chinese Academy of Sciences.

\section*{Data Availability}

The LAMOST data employed in this article are available after September
2022 to the users out of China for download from LAMOST DR8, at \url{http://www.lamost.org/dr8/}. And the software for this pipeline, trained model, training set, test data sets and the produced catalog are available at \url{https://github.com/xrli/LASSO-MLPNet}.


\section*{Footnotes}

software: Numpy \citep{harris2020array}, Scipy \citep{virtanen2020scipy}, Astropy \citep{price2018astropy}, Matplotlib \citep{hunter2007matplotlib}, Scikit-learn\citep{pedregosa2011scikit}.



\bibliographystyle{mnras}
\bibliography{example} 

\begin{thebibliography}{}
\makeatletter
\relax
\def\mn@urlcharsother{\let\do\@makeother \do\$\do\&\do\#\do\^\do\_\do\%\do\~}
\def\mn@doi{\begingroup\mn@urlcharsother \@ifnextchar [ {\mn@doi@}
  {\mn@doi@[]}}
\def\mn@doi@[#1]#2{\def\@tempa{#1}\ifx\@tempa\@empty \href
  {http://dx.doi.org/#2} {doi:#2}\else \href {http://dx.doi.org/#2} {#1}\fi
  \endgroup}
\def\mn@eprint#1#2{\mn@eprint@#1:#2::\@nil}
\def\mn@eprint@arXiv#1{\href {http://arxiv.org/abs/#1} {{\tt arXiv:#1}}}
\def\mn@eprint@dblp#1{\href {http://dblp.uni-trier.de/rec/bibtex/#1.xml}
  {dblp:#1}}
\def\mn@eprint@#1:#2:#3:#4\@nil{\def\@tempa {#1}\def\@tempb {#2}\def\@tempc
  {#3}\ifx \@tempc \@empty \let \@tempc \@tempb \let \@tempb \@tempa \fi \ifx
  \@tempb \@empty \def\@tempb {arXiv}\fi \@ifundefined
  {mn@eprint@\@tempb}{\@tempb:\@tempc}{\expandafter \expandafter \csname
  mn@eprint@\@tempb\endcsname \expandafter{\@tempc}}}

\bibitem[\protect\citeauthoryear{Bailer-Jones, Irwin, Gilmore  \& von
  Hippel}{Bailer-Jones et~al.}{1997}]{bailer1997physical}
Bailer-Jones C.~A.,  Irwin M.,  Gilmore G.,   von Hippel T.,  1997, Monthly
  Notices of the Royal Astronomical Society, 292, 157

\bibitem[\protect\citeauthoryear{Choi, Dotter, Conroy, Cantiello, Paxton  \&
  Johnson}{Choi et~al.}{2016}]{choi2016mesa}
Choi J.,  Dotter A.,  Conroy C.,  Cantiello M.,  Paxton B.,   Johnson B.~D.,
  2016, The Astrophysical Journal, 823, 102

\bibitem[\protect\citeauthoryear{Cui et~al.,}{Cui et~al.}{2012}]{cui2012}
Cui X.-Q.,  et~al., 2012, Research in Astronomy and Astrophysics, 12, 1197

\bibitem[\protect\citeauthoryear{Dotter}{Dotter}{2016}]{dotter2016mesa}
Dotter A.,  2016, The Astrophysical Journal Supplement Series, 222, 8

\bibitem[\protect\citeauthoryear{Gal \& Ghahramani}{Gal \&
  Ghahramani}{2016}]{gal2016dropout}
Gal Y.,  Ghahramani Z.,  2016, in international conference on machine learning.
  pp 1050--1059

\bibitem[\protect\citeauthoryear{Harris et~al.,}{Harris
  et~al.}{2020}]{harris2020array}
Harris C.~R.,  et~al., 2020, Nature, 585, 357

\bibitem[\protect\citeauthoryear{Holtzman et~al.,}{Holtzman
  et~al.}{2018}]{holtzman2018}
Holtzman J.~A.,  et~al., 2018, The Astronomical Journal, 156, 125

\bibitem[\protect\citeauthoryear{Hron, Matthews  \& Ghahramani}{Hron
  et~al.}{2018}]{hron2018variational}
Hron J.,  Matthews A.,   Ghahramani Z.,  2018, in International Conference on
  Machine Learning. pp 2019--2028

\bibitem[\protect\citeauthoryear{Hunter}{Hunter}{2007}]{hunter2007matplotlib}
Hunter J.~D.,  2007, Computing in science \& engineering, 9, 90

\bibitem[\protect\citeauthoryear{J{\"o}nsson et~al.,}{J{\"o}nsson
  et~al.}{2018}]{jonsson2018apogee}
J{\"o}nsson H.,  et~al., 2018, The Astronomical Journal, 156, 126

\bibitem[\protect\citeauthoryear{Koleva, Prugniel, Bouchard  \& Wu}{Koleva
  et~al.}{2009}]{koleva2009ulyss}
Koleva M.,  Prugniel P.,  Bouchard A.,   Wu Y.,  2009, Astronomy \&
  Astrophysics, 501, 1269

\bibitem[\protect\citeauthoryear{Leung \& Bovy}{Leung \&
  Bovy}{2019}]{leung2019deep}
Leung H.~W.,  Bovy J.,  2019, Monthly Notices of the Royal Astronomical
  Society, 483, 3255

\bibitem[\protect\citeauthoryear{Li, Wu, Luo, Zhao, Lu, Zuo, Yang  \& Wang}{Li
  et~al.}{2014}]{li2014sdss}
Li X.,  Wu Q.~J.,  Luo A.,  Zhao Y.,  Lu Y.,  Zuo F.,  Yang T.,   Wang Y.,
  2014, The Astrophysical Journal, 790, 105

\bibitem[\protect\citeauthoryear{Li, Lu, Comte, Luo, Zhao  \& Wang}{Li
  et~al.}{2015}]{li2015linearly}
Li X.,  Lu Y.,  Comte G.,  Luo A.,  Zhao Y.,   Wang Y.,  2015, The
  Astrophysical Journal Supplement Series, 218, 3

\bibitem[\protect\citeauthoryear{Liu et~al.,}{Liu et~al.}{2014}]{liu2014}
Liu C.,  et~al., 2014, The Astrophysical Journal, 790, 110

\bibitem[\protect\citeauthoryear{Liu, Zhao  \& Hou}{Liu et~al.}{2015}]{liu2015}
Liu X.-W.,  Zhao G.,   Hou J.-L.,  2015, Research in Astronomy and
  Astrophysics, 15, 1089

\bibitem[\protect\citeauthoryear{Luo et~al.,}{Luo et~al.}{2015}]{luo2015}
Luo A.-L.,  et~al., 2015, Research in Astronomy and Astrophysics, 15, 1095

\bibitem[\protect\citeauthoryear{Majewski et~al.,}{Majewski
  et~al.}{2017}]{majewski2017}
Majewski S.~R.,  et~al., 2017, The Astronomical Journal, 154, 94

\bibitem[\protect\citeauthoryear{Masseron et~al.,}{Masseron
  et~al.}{2014}]{masseron2014ch}
Masseron T.,  et~al., 2014, Astronomy \& Astrophysics, 571, A47

\bibitem[\protect\citeauthoryear{Pedregosa et~al.,}{Pedregosa
  et~al.}{2011}]{pedregosa2011scikit}
Pedregosa F.,  et~al., 2011, the Journal of machine Learning research, 12, 2825

\bibitem[\protect\citeauthoryear{P{\'e}rez et~al.,}{P{\'e}rez
  et~al.}{2016}]{perez2016}
P{\'e}rez A. E.~G.,  et~al., 2016, The Astronomical Journal, 151, 144

\bibitem[\protect\citeauthoryear{Plez, Brett  \& Nordlund}{Plez
  et~al.}{1992}]{plez1992spherical}
Plez B.,  Brett J.~M.,   Nordlund A.,  1992, Astronomy and Astrophysics, 256,
  551

\bibitem[\protect\citeauthoryear{Price-Whelan et~al.,}{Price-Whelan
  et~al.}{2018}]{price2018astropy}
Price-Whelan A.~M.,  et~al., 2018, The Astronomical Journal, 156, 123

\bibitem[\protect\citeauthoryear{Prugniel \& Soubiran}{Prugniel \&
  Soubiran}{2001}]{prugniel2001database}
Prugniel P.,  Soubiran C.,  2001, Astronomy \& Astrophysics, 369, 1048

\bibitem[\protect\citeauthoryear{Prugniel, Soubiran, Koleva  \&
  Borgne}{Prugniel et~al.}{2007}]{prugniel2007new}
Prugniel P.,  Soubiran C.,  Koleva M.,   Borgne D.~L.,  2007, arXiv preprint
  astro-ph/0703658

\bibitem[\protect\citeauthoryear{Sch{\"o}lkopf, Smola  \&
  M{\"u}ller}{Sch{\"o}lkopf et~al.}{1997}]{scholkopf1997kernel}
Sch{\"o}lkopf B.,  Smola A.,   M{\"u}ller K.-R.,  1997, in International
  conference on artificial neural networks. pp 583--588

\bibitem[\protect\citeauthoryear{Taylor}{Taylor}{2017}]{taylor2017topcat}
Taylor M.,  2017, arXiv preprint arXiv:1711.01885

\bibitem[\protect\citeauthoryear{Tibshirani}{Tibshirani}{1996}]{tibshirani1996regression}
Tibshirani R.,  1996, Journal of the Royal Statistical Society: Series B
  (Methodological), 58, 267

\bibitem[\protect\citeauthoryear{Ting, Rix, Conroy, Ho  \& Lin}{Ting
  et~al.}{2017}]{ting2017measuring}
Ting Y.-S.,  Rix H.-W.,  Conroy C.,  Ho A.~Y.,   Lin J.,  2017, The
  Astrophysical Journal Letters, 849, L9

\bibitem[\protect\citeauthoryear{Ting, Conroy, Rix  \& Cargile}{Ting
  et~al.}{2019}]{ting2019payne}
Ting Y.-S.,  Conroy C.,  Rix H.-W.,   Cargile P.,  2019, The Astrophysical
  Journal, 879, 69

\bibitem[\protect\citeauthoryear{Virtanen et~al.,}{Virtanen
  et~al.}{2020}]{virtanen2020scipy}
Virtanen P.,  et~al., 2020, Nature methods, 17, 261

\bibitem[\protect\citeauthoryear{Wang et~al.,}{Wang
  et~al.}{2020}]{wang2020spcanet}
Wang R.,  et~al., 2020, The Astrophysical Journal, 891, 23

\bibitem[\protect\citeauthoryear{Wu et~al.,}{Wu et~al.}{2011}]{wu2011automatic}
Wu Y.,  et~al., 2011, Research in Astronomy and Astrophysics, 11, 924

\bibitem[\protect\citeauthoryear{Wu, Du, Luo, Zhao  \& Yuan}{Wu
  et~al.}{2014}]{wu2014automatic}
Wu Y.,  Du B.,  Luo A.,  Zhao Y.,   Yuan H.,  2014, Proceedings of the
  International Astronomical Union, 10, 340

\bibitem[\protect\citeauthoryear{Xiang et~al.,}{Xiang
  et~al.}{2017}]{xiang2017estimating}
Xiang M.-S.,  et~al., 2017, Monthly Notices of the Royal Astronomical Society,
  464, 3657

\bibitem[\protect\citeauthoryear{Xiang et~al.,}{Xiang
  et~al.}{2019}]{xiang2019abundance}
Xiang M.,  et~al., 2019, The Astrophysical Journal Supplement Series, 245, 34

\bibitem[\protect\citeauthoryear{Zhang, Liu  \& Deng}{Zhang
  et~al.}{2020}]{zhang2020deriving}
Zhang B.,  Liu C.,   Deng L.-C.,  2020, The Astrophysical Journal Supplement
  Series, 246, 9

\bibitem[\protect\citeauthoryear{Zhao, Zhao, Chu, Jing  \& Deng}{Zhao
  et~al.}{2012}]{zhao2012lamost}
Zhao G.,  Zhao Y.-H.,  Chu Y.-Q.,  Jing Y.-P.,   Deng L.-C.,  2012, Research in
  Astronomy and Astrophysics, 12, 723

\makeatother
\end{thebibliography}







\bsp	
\label{lastpage}
\end{document}